\documentclass[onecolumn]{emulateapj}

\usepackage{epstopdf}
\usepackage{ulem}
\usepackage{amssymb}
\usepackage{natbib}
\usepackage{times}
\usepackage{graphicx,bm,amssymb}
\usepackage{pstricks}
\usepackage{psfrag}
\usepackage[colorlinks=true,linkcolor=blue,citecolor=blue]{hyperref}
\newcommand{\be}{\begin{equation}}
\newcommand{\ee}{\end{equation}}
\newcommand{\bse}{\begin{subequations}}
\newcommand{\ese}{\end{subequations}}
\newcommand{\bary}{\begin{eqnarray}}
\newcommand{\eary}{\end{eqnarray}}
\def\aj{AJ}
\def\apj{ApJ}
\def\apjl{ApJ}
\def\apjs{ApJS}
\def\aap{A\&A}
\def\jcap{J. Cosmology Astropart. Phys.}
\def\mnras{MNRAS}
\def\pasa{PASA}
\def\prd{Phys.~Rev.~D}
\def\rmxaa{Rev. Mexicana Astron. Astrofis.}

\shorttitle{Hypercritical accretion scenario in RX J0822-4300}
\shortauthors{Fraija N. et al.}

\begin{document}
\title{Could an hypercritical accretion be associated with\\ the atypical magnetic-field behavior in RX J0822-4300?}
             
\author{N. Fraija$^{1\dagger}$, C.~G. Bernal$^2$,  G. Morales$^1$ and  R.~Negreiros$^{3}$} 
\affil{$^1$ Instituto de Astronom\' ia, Universidad Nacional Aut\'onoma de M\'exico, Circuito Exterior, C.U., A. Postal 70-264, 04510 M\'exico D.F., M\'exico.\\
$^2$ Instituto de Matem\'atica, Estat\'istica e F\'isica, Universidade Federal de Rio Grande,  Av. Italia km 8 Bairro Carreiros, Rio Grande, RS, Brazil \\
$^3$ Instituto de F\'isica, Universidade Federal Fluminense, Av. Gal. Milton Tavares de Souza s/n, Gragoata, Niteroi, 24210-346, Brazil}
     
\email{$\dagger$nifraija@astro.unam.mx}            


\begin{abstract}
Recent X-ray observations in the central compact remnant of Puppis A and other young pulsars have provided convincing evidence about the anti-magnetar nature of the so-called central compact objects (CCOs). The measured period derivative, corrected by the proper movement, provides exceptionally low superficial magnetic fields for such sources. Using the dipole radiation canonical model,  the superficial magnetic field of the CCO (RX J0822-4300) in Puppis A was estimated to be $B\simeq 2.9 \times 10^{10}$ G. In this paper,  we present a numerical model to interpret the unusual  magnetic-field behavior in RX J0822-4300.  Using the magnetohydrodynamic simulations during the hypercritical accretion phase,  we propose that a variety of MeV neutrinos is created given evidence about the submergence of the magnetic field in the  pulsar.   We investigate the production, propagation, oscillations, and detection of  MeV neutrinos from this source. The detection of neutrino events with their flavor ratios would be a compelling signature of the decreasing evolution of the magnetic field not only in RX J0822-4300 but also in CCO candidates.\\ 
\end{abstract}

\keywords{supernovae: individual: Puppis A -- pulsars: individual: RX J0822-4300 -- neutrino: cooling -- neutrino: oscillations -- stars: neutron  --  accretion  -- hydrodynamics -- magnetic field}

\section{Introduction}\label{sec-Intro}
Central Compact Objects (CCOs) are neutron stars (NSs) located close to the geometrical central regions of young (0.3 - 7 kyr) supernova remnants  \citep[SNRs,][]{2010PNAS..107.7147K}.  These sources are characterized by emitting photons in the X-ray band with typical luminosities in the range $L_X\sim 10^{33} - 10^{34}\,  {\rm erg\, s^{-1}}$ \citep{2004IAUS..218..239P}.  Their spectra are thermal-like,  described in general by two black-body (BB) functions with high temperatures  (ranging from 0.2 to 0.6 keV) and small emitting radii ($\sim$ 0.1 - 5 km). The origin of the magnetic fields within NSs is still an unsolved problem.  Estimations of the superficial magnetic field (B-field) show that it is small, favoring the hidden B-field scenario. In this scenario, the B-field is immersed by the hypercritical accretion material (see e.g., \cite{1995ApJ...440L..77M} and \cite{1999A26A...345..847G}, \cite{Bernal2013} and references therein).\\
\\
Following the arguments of  \cite{1995ApJ...440L..77M},  simple 1D ideal magnetohydrodynamic (MHD) simulations were performed by \cite{1999A26A...345..847G}  to show the effect of the  hypercritical accretion over the B-field in a newborn NS.   These simulations displayed a rapid submergence of the B-field into the NS \citep{1999A26A...345..847G}.   This  scenario was revisited by several authors in order to study the evolution of the B-field in these compact objects.   \cite{2015MNRAS.451..455F} and \cite{2016MNRAS.462.3646B} showed that thermal neutrinos created during this phase experiment resonant conversions,  playing an outstanding role. They estimated from Cassiopeia A and Kesteven 79  the number of neutrino events and the flavor ratio expected on Earth.\\
\\
Puppis A, located at $2.2\pm 0.3$ kpc \citep{1995AJ....110..318R}, is one of the most interesting SNRs in the Southern hemisphere.  Its age is ranging from 3700 and 5200 yrs; 3700 yrs when the proper motion of optical filaments is considered, and 5200 yrs when the projected distance from its explosion center is taken into account.   
The High-Resolution Imager on Einstein observatory \citep[HRI: see][]{1979ApJ...230..540G} discovered the X-ray compact object RX J0822-4300, in Puppis A.    One decade later, ROSAT observations confirmed that RX J0822-4300 was a central compact stellar remnant formed in the SN event \citep{1996ApJ...465L..43P}.    Based on X-ray observations performed by Chandra and XMM-Newton, \cite{2006A&A...454..543H} showed  that the X-ray emission was consistent with a two BB model with temperatures $T_1=(2.35-2.91)\times 10^6\,\,{\rm K}$ and $T_2=(4.84-5.3)\times 10^6\,\, {\rm K}$. It is worth noting  that this object has not been detected either radio (as a radio pulsar) or optical (with magnitude 25$\lesssim$ B-band and 23.6 $\lesssim$ R-band) frequencies. These observations together with the lack of long-term flux variations confirmed to this source RX J0822-4300 as a NS candidate with a low B-field \citep{2006A&A...454..543H}. Later, \cite{2013ApJ...765...58G} could infer the strength of the B-field of $2.9 \times 10^{10}$ G using  a dipole model with values of P = 112 ms and ${\rm \dot{P}}=(9.28 \pm 0.36) \times 10^{-18}$ \citep{2013ApJ...765...58G}.  \\
\\
Taking in consideration that \cite{Shabaltas2012} and \cite{Popov2015} argued that  the hidden B-field scenario could be applicable to Puppis A, in the current work, we use the FLASH Code in order to simulate the hyper-accretion phase, the B-field evolution close to the stellar surface and the neutrino luminosity from RX J0822-4300.  Additionally,  we investigate the production, propagation, oscillation, and detection of  thermal neutrinos from this source.   The paper is arranged as follows. In Section \ref{sec-Physics} we briefly describe the numerical model  of the hypercritical phase. In Section \ref{sec-Neutrinos} we study the production, propagation, oscillation, and detection of  thermal neutrinos.   In Section \ref{sec-discussion} we discuss our results and finally, the conclusions are presented in Section \ref{sec-Results}
%
\section{Numerical Model}\label{sec-Physics}
We analyze the hypercritical accretion phase onto the newborn NS (RX J0822-4300) inside Puppis A, using the numerical model previously shown in \cite{Bernal2010,Bernal2013}.  We are interested in following the evolution of the B-field and  the production, propagation and oscillations of  thermal neutrinos during the hypercritical accretion phase.   Here, we extend the scenario presented in \cite{Bernal2010,Bernal2013} and focus on the neutrino analysis, the dynamics and morphology of the magnetized and thermal plasma around  RX J0822-4300.  Due to the similarity of Puppis A with Kes 79 \citep{2013MNRAS.434..123V}, in the current paper we adopt the standard parameters reported for the NS in Puppis A: $M \simeq 1.45$ $M_{\odot}$, $R \simeq10$ km, and $B \simeq 10^{12}$ G,  which correspond to progenitor models of pre-supernova in the range of  25 - 35 $M_{\odot}$ \citep[see, ][and references therein.]{Chevalier2005, Shabaltas2012}\\
\\
In order to tackle numerically the hypercritical accretion phase, a customized version of the Eulerian and multi-physics parallel FLASH Code is used \citep{Fryxell2000}.   The complete set of MHD equations are solved by the FLASH Code using the unsplit staggered mesh algorithm which is included in the more recent version of FLASH Code\footnote{http://flash.uchicago.edu/site/}.   The unsplit staggered mesh algorithm uses a directionally unsplit scheme to evolve the MHD governing equations. A more detailed description of the Unsplit Staggered Mesh Scheme can be found in \cite{LeeDongwook2013}.   We use the Helmholtz Unit that includes contributions coming from radiation, ionized nuclei, and degenerate/relativistic electrons and positrons. Although  the Helmholtz EoS has been widely used in several astrophysical environments,  we have used it with caution. Full details of the Helmholtz algorithms are provided in \cite{Timmes2000}. The neutrino cooling processes and the Equation of State (EoS) algorithm  are also included in this code through a customized Unit. \\ 
We carry out 2-D MHD simulations of the hypercritical phase using computational domain columns of $(20 \times 100)$ km. The complete procedure,  the initial configuration for the thermodynamic variables and the lateral boundary conditions are explicitly described in \cite{2016MNRAS.462.3646B}.
\section{Neutrinos}\label{sec-Neutrinos}
Neutrinos represent an active role in the SN dynamics, from  core collapse phenomena up to the catastrophic explosion mechanisms. In the initial stage of  the core collapse,  densities become larger than $\gtrsim 10^{9-10}\,{\rm g/cm^3}$ and electrons are captured by nuclei with $\nu_e$ freely escaping from the collapsing star. When the core density is too high so that neutrinos are trapped, the chemical potential will be in equilibrium, allowing the diffusive emission of all neutrino flavors. More than 95 \% of the gravitational binding energy will  be  released as neutrinos, as shown by simulations \citep[see i.e.][]{2012MNRAS.426.1940K, 2016MNRAS.459.4174G}.\\
\\
On the NS surface ({\small $r_{ns}=10^{6}\,\,{\rm cm}$}),  a new crust of hundreds of meters endowed with a strong  B-field in the range $10^{11}\leq {\bf B}\leq10^{13}\,{\rm G}$ is created and later submerged by the hypercritical accretion phase. Photons and thermal neutrinos produced and  confined in this region are thermalized to a few MeV.  In this phase,  energy losses become dominated primarily by the annihilation processes, involving the formation of  $\nu_x {\bar\nu}_x$ pairs when  $e^\pm$ pairs are annihilated close to the stellar surface.  Other important neutrino processes involved in the customized routine are  $e^-e^+$ pair annihilations  ($e^-+e^+\to \nu_{\rm x} + \bar{\nu}_{\rm x}$), plasmon decays  ($\gamma\to \nu_{\rm x} + \bar{\nu}_{\rm x}$) and photo-neutrino emission  ($\gamma+e^-\to e^- + \nu_{\rm x} + \bar{\nu}_{\rm x}$).  The subscript ${\rm x}$ indicates the neutrino flavor: electron, muon, or tau.\\
\subsection{Effective Potential}
In the hypercritical accretion phase, thermal neutrinos are generated in the region I and consequently, they will oscillate in their paths  in accordance with the neutrino effective potential of each region. 
\subsubsection{Region I: New crust of NS surface {\small ($ r\leq [r_{\rm ns}+r_{\rm c}]$)}}
\paragraph{Moderate B-field limit}
In the moderate B-field approximation,  the neutrino effective potential is \citep{2014ApJ...787..140F, 2015MNRAS.451..455F}
\begin{eqnarray}\label{Veffm}
V_{\rm eff,I(m)}=V_o\left[\sum^{\infty}_{l=0}(-1)^l\sinh\alpha_l   \left[F_m-G_m\cos\varphi \right] -4\frac{m^2_e}{m^2_W}\,\frac{E_\nu}{m_e}\sum^\infty_{l=0}(-1)^l\cosh\alpha_l  \left[J_m-H_m\cos\varphi \right]  \right]\,,
 \end{eqnarray}
where $V_o=\frac{\sqrt2\,G_F\,m_e^3 B}{\pi^2\,B_c}$,  $B_c=4.414\times 10^{13}\,{\rm G}$ is the critical B-field,  $m_e$ and $m_W$ are the electron and boson masses, respectively,  $G_F$ is the Fermi constant,  $\alpha_l=\beta\mu(l+1)$ and  
\bary
F_{\rm m}&=&\biggl(1+2\frac{E^2_\nu}{m^2_W}\biggr)K_1(\sigma_l)+2\sum^\infty_{n=1}\lambda_n\biggl(1+\frac{E^2_\nu}{m^2_W}  \biggr)K_1(\sigma_l\lambda_n)\,,\nonumber\\
G_{\rm m}&=&\biggl(1-2\frac{E^2_\nu}{m^2_W}\biggr)K_1(\sigma_l)-2\sum^\infty_{n=1}\lambda_n\frac{E^2_\nu}{m^2_W} K_1(\sigma_l\lambda_n)\,,\nonumber\\
J_{\rm m}&=& \frac34 K_0(\sigma_l)+\frac{K_1(\sigma_l)}{\sigma_l}+\sum^\infty_{n=1}\lambda^2_n\biggl[K_0(\sigma_l\lambda)+\frac{K_1(\sigma_l\lambda)}{\sigma_l\lambda}-\frac{K_0(\sigma_l\lambda)}{2\lambda^2_n}  \biggr]\,,\nonumber\\
H_{\rm m}&=& \frac{K_1(\sigma_l)}{\sigma_l}+ \sum^\infty_{n=1}\lambda^2_n   \biggl[\frac{K_1(\sigma_l\lambda)}{\sigma_l\lambda} - \frac{K_0(\sigma_l\lambda)}{2\lambda^2_n}  \biggr]\,,
\eary
with $\lambda^2_n=1+2\,n\,B/B_c $, K$_i$ the modified Bessel function of integral order i and $\sigma_l=\beta m_e(l+1)$.\\
\paragraph{Weak B-field limit}
In the weak B-field approximation, the effective potential does not depend on the Landau levels.  Therefore,  it can be written as  \citep{2014ApJ...787..140F}
\begin{eqnarray}\label{Veffw}
V_{\rm eff,I(w)}=V_o\left[\sum^{\infty}_{l=0}(-1)^l\sinh\alpha_l  \left[F_w-G_w\cos\varphi \right]-4\frac{m^2_e}{m^2_W}\,\frac{E_\nu}{m_e}\sum^\infty_{l=0}(-1)^l\cosh\alpha_l \left[J_w-H_w\cos\varphi \right]\right]\,, 
\end{eqnarray}
where in this case the functions F$_{\rm w}$, G$_{\rm w}$, J$_{\rm w}$ and H$_{\rm w}$ are 
\bary
F_{\rm w}&=&\biggl(2+2\frac{E^2_\nu}{m^2_W}\biggr) \biggl(\frac{K_0(\sigma_l)}{\sigma_l}+2\frac{K_1(\sigma_l)}{\sigma_l^2} \biggr) \frac{B_c}{B}-K_1(\sigma_l)\,,\nonumber\\
G_{\rm w}&=&K_1(\sigma_l)-\frac{2B_c}{B}\frac{E^2_\nu}{m^2_W}    \biggl(\frac{K_0(\sigma_l)}{\sigma_l}+2\frac{K_1(\sigma_l)}{\sigma_l^2}\biggr)\,, \nonumber\\
J_{\rm w}&=&\biggl(\frac12+\frac{3B_c}{B\,\sigma_l^2}\biggr)K_0(\sigma_l)+\frac{B_c}{B}  \biggl(1+\frac{6}{\sigma_l^2}\biggr)     \frac{K_1(\sigma_l)}{\sigma_l}\,,\nonumber\\
H_{\rm w}&=& \biggl(\frac12+\frac{B_c}{B\,\sigma_l^2} \biggr)K_0(\sigma_l)+\frac{B}{B_c} \biggl(\frac{2}{\sigma_l^2}-\frac12\biggr)\frac{K_1(\sigma_l)}{\sigma_l}\,,\nonumber\\
\eary
respectively.\\
\paragraph{Only Neutrino Background}
In a newly born NS, the neutrinos with a mean free path less than the depth of the medium are trapped in a region called the neutrino-sphere. In these neutrino-spheres, neutrinos and anti-neutrinos have different average energies  $\langle E_{\nu_e}\rangle\simeq10\,\,{\rm MeV}$, $\langle \bar{E}_{\nu_e}\rangle\simeq15\,\,{\rm MeV}$ and $\langle \bar{E}_{\nu_{\mu/\tau}}\rangle=\langle E_{\nu_{\mu/\tau}}\rangle \simeq 20\,\,{\rm MeV}$  \citep{2007fnpa.book.....G}.\\
If the medium contains only the neutrinos and anti-neutrinos of all flavors, then the neutrino effective potential  is \citep{2009JCAP...11..024S}
\bary
V_{\rm eff, I(bk)}&=&\sqrt2 G_F\biggl[(N_{\nu_e}-\bar{N}_{\nu_e})-(N_{\nu_\mu}-\bar{N}_{\nu_\mu})\nonumber\\
&&-\frac83\frac{1}{M_Z^2}\biggl\{\langle E_{\nu_e}\rangle\biggl(\langle E_{\nu_e}\rangle
N_{\nu_e}+\langle \bar{E}_{\nu_e}\rangle \bar{N}_{\nu_e}\biggr)
-\biggl(\langle E_{\nu_\mu}\rangle^2 N_{\nu_\mu}+\langle {E}_{\nu_\mu}\rangle^2 \bar{N}_{\nu_\mu}\biggr)\biggr\}\biggr].
\eary
By assuming that the number density of neutrinos ($N_{\nu}$) and anti-neutrinos  ($\bar{N}_{\nu}$) of all flavors is the same inside the neutrino-sphere, i.e. $N_{\nu_{\rm x}}=\bar{N}_{\nu_{\rm x}}$, the effective neutrino potential becomes
\be
V_{\rm eff, I(bk)}=2.2\times 10^{-14}\, {\rm eV}\,.
\ee
\subsubsection{Regions II: Quasi-hydrostatic envelope {\small ($[r_{\rm ns}+r_{\rm c}]\leq r\leq r_{\rm s}$)} }
In the quasi-hydrostatic envelope, the reverse shock gives rise to the hypercritical accretion onto the new NS surface through the fall-back stage.  Later, a new expansive shock is formed by the material accreted and bounced off against the new NS surface.  The expansive shock  leads  an  envelope  in  quasi-hydrostatic  equilibrium   with free-fall material falling over it.  Taking into account the typical values $M=1.4\,M_{\odot}$ and  $\dot{m}=500 M_{\odot}\, {\rm yr}^{-1}$, the shock radius is {\small $r_s\simeq 7.7\times 10^8\,{\rm cm} \left(M/1.4M_\odot \right)^{-0.04}\,\left(r_{\rm ns}/10^6 {\rm cm}\right)^{1.48}\\ \left(\dot{m}/500\, M_\odot\, {\rm yr}^{-1} \right)^{-0.37}$}. From the strong shock conditions, the density in this region is $\rho(r)= 7.6\times 10^2\,\left(\frac{r_s}{r} \right)^3\,{\rm g\, cm^{-3}}$ and the effective potential is
\be
V_{\rm eff, II}= 7.6\times 10^{2}\,V_a\,  \left(\frac{r_s}{r} \right)^3\,,  
\ee
where $V_a=\sqrt2 G_F\, N_A \,Y_e$ with $Y_e$ the number of electron per nucleon and  $N_A$ the Avogadro's number.
\subsubsection{Regions III: Free-fall {\small ($r_s\leq r\leq r_h$)}}
In the free-fall region, the material begins falling with a speed $v(r)=\sqrt{\frac{2GM}{r}}$ and density $\rho(r)=\frac{\dot{m}}{4\pi r^2 v(r)}$.  Given the typical values $M=1.4\,M_{\odot}$ and  $\dot{m}=500 M_{\odot}\, {\rm yr}^{-1}$,  the density of the free-fall material is {\small $\rho(r)=5.7\times 10^{-2} \left(\frac{r_h}{r} \right)^{\frac32}{\rm g\, cm^{-3}}$} and the effective potential is
\be
V_{\rm eff, III}=5.7\times 10^{-2}\,V_a\,\left(\frac{r_h}{r} \right)^{\frac32}\,,
\ee
where {\small $r_h= 10^{10.8}\,{\rm cm}$}.

\subsubsection{Regions IV: {\small ($r_h\leq r$)}}
Taking into account the typical profile for a pre-supernova, the density of the outer layer could be given by 
\be
V_{\rm eff, IV}=10^{-5.2}\,V_a\,A\left(\frac{R_\star}{r} - 1\right)^k\,,
\ee
with
{\small
\bary
 (k,A)=\cases{
{\small (2.1,20)}; \hspace{0.5cm}  r_h< r < r_a, \cr 
{\small (2.5,1)};\,\hspace{0.6cm}   r>r_a\,,\cr
}
\eary
}
{\small $\,\,R_{\star}\simeq10^{12.5}\,{\rm cm}$} and   {\small $r_a= 10^{11}\,{\rm cm}$}.  
\subsection{Neutrino Oscillation}
Solar, atmospheric and accelerator neutrino experiments  have exhibited through the masses and mixing parameters convincing evidences about neutrino oscillations.    Table \ref{table:neu_par} shows the neutrino oscillation parameters from global analysis considering new data from  T2K \citep{2017arXiv170406409A,2017arXiv170100432T}, NO$\nu$A \citep{2017arXiv170105891T,2017arXiv170303328T},  RENO \citep{2017arXiv171008204S, 2018arXiv180104049P},   Double Chooz \citep{2012PhRvD..86e2008A}, IceCube DeepCore \citep{2014arXiv1410.7227I}, ANTARES \citep{2012PhLB..714..224A} and Super-Kamiokande \citep{2018PhRvD..97g2001A} neutrino experiments. The parameter values of solar, atmospheric and accelerator neutrino experiments correspond to  $\Delta m^2_{21}\equiv m_2^2-m_1^2=\Delta m^2_{\rm sol}$,  $\Delta m^2_{32}\equiv m_3^2-m_2^2=\Delta m^2_{\rm atm}$ and $\Delta m^2_{31}\equiv m_3^2-m_1^2=\Delta m^2_{\rm acc}$.  In this paper, the natural units $\hbar=c=k_B=1$  are used.\\
\begin{table*}[h!]
\begin{center}\renewcommand{\arraystretch}{2.}\addtolength{\tabcolsep}{5pt}
\caption{Neutrino oscillation parameters from global analysis \citep{2017arXiv170801186D}}\label{table:neu_par}
\begin{tabular}{ c c c }
\hline \hline
Parameter &      Best fit (NH) \\
\hline\hline
$\frac{\sin^2\theta_{12}}{10^{-1}}$  & $3.20^{+0.20}_{-0.16}$ \\
$\theta_{12}$  &    $34^\circ.5^{+1.2}_{-1.0}$ \\
$\frac{\sin^2\theta_{23}}{10^{-1}}$  & $5.47^{+0.20}_{-0.30}$    \\
$\theta_{23}$  &   $47^\circ.7^{+1.2}_{-1.7}$  \\
$\frac{\sin^2\theta_{13}}{10^{-2}}$  &$2.160^{+0.083}_{-0.069}$  \\
$\theta_{13}$  & $8^\circ.45^{+0.16}_{-0.14}$    \\
$\frac{\Delta m^2_{21}}{10^{-5}\,{\rm eV^2}}$ & $7.55^{+0.20}_{-0.16}$  \\
$\frac{\Delta m^2_{32}}{10^{-3}\,{\rm eV^2}}$ & $2.457^{+0.047}_{-0.047}$   \\
$\frac{\Delta m^2_{31}}{10^{-3}\,{\rm eV^2}}$ &  $2.50^{+0.03}_{-0.03}$ \\
\hline
\hline
 \end{tabular}
\end{center}
\end{table*}
\subsubsection{In Vacuum}
Neutrino oscillates in vacuum between the compact object RX J0822-4300 and Earth. The matrix of transition for neutrinos created in RX J0822-4300 can be written as \citep{2016MNRAS.462.3646B}
{\small
\be
{\pmatrix
{
\nu_e   \cr
\nu_\mu   \cr
\nu_\tau   \cr
}_{E}}
=
{\pmatrix
{
0.534	  & 0.266	  & 0.200\cr
 0.266	  & 0.366	  &  0.369\cr
 0.200	  & 0.368	  & 0.432\cr
}}
{\pmatrix
{
\nu_e   \cr
\nu_\mu   \cr
\nu_\tau   \cr
}_{RX}}
\label{matrixosc}\,.
\ee
}
\subsubsection{In Matter}
\paragraph{Two-Neutrino Mixing}
The evolution equation of the neutrino propagation in matter for two-flavor mixing is given by
{\small
\be
i
{\pmatrix {\dot{\nu}_{e} \cr \dot{\nu}_{\mu}\cr}}
={\pmatrix
{V_{\rm eff}-\frac{\Delta m^2}{2E_{\nu}} \cos 2\theta & \frac{\Delta m^2}{4E_{\nu}}\sin 2\theta\,, \cr
\frac{\Delta m^2}{4E_{\nu}}\sin 2\theta  & 0\cr}}\,
{\pmatrix
{\nu_{e} \cr \nu_{\mu}\cr}},
\ee
}

where $V_{\rm eff}$ is the effective neutrino potential and $E_{\nu}$ is the neutrino energy.  The neutrino oscillation process $\nu_e\leftrightarrow \nu_{\mu, \tau}$ has been considered. The transition probability in this case is
{\small
\be
P_{\nu_e\rightarrow {\nu_{\mu}{(\nu_\tau)}}}(t) =  \left[ \frac{\Delta m^2}{2\,{\rm \omega\,E_{\nu}}}\sin\left (\frac{\omega t}{2}\right)\right]^2\,\,\,\,\,{\rm with}\,\,\,\,\,\omega=\sqrt{\left(V_{eff}-\frac{\Delta m^2}{2E_{\nu}} \cos 2\theta\right)^2+  \left(\frac{\Delta m^2}{2\,E_{\nu}} \sin 2\theta\right)^2}\,.
\label{prob}
\ee
}
From the resonance condition, the effective potential and the oscillation length become
{\small
\be
V_{\rm eff} -  \frac{\Delta m^2}{2E_{\nu}} \cos 2\theta = 0\,\,\,\,\,{\rm and}\,\,\,\,\,L_{res}=\frac{4\pi E_\nu}{\Delta m^2\sin 2\theta}\,,
\label{reso}
\ee
}
respectively. Details are shown in \cite{2014MNRAS.437.2187F}.
\paragraph{Three-Neutrino Mixing}
The evolution equation of the neutrino propagation in matter for three-flavor mixing is given by
\be
i\frac{d}{dt} 
{\pmatrix
{
\nu_e   \cr
\nu_\mu   \cr
\nu_\tau   \cr
}}=
\mathcal{H}
{\pmatrix
{
\nu_e   \cr
\nu_\mu   \cr
\nu_\tau   \cr
}}\,,
\ee
where the effective Hamiltonian is
\be
\mathcal{H}=U\cdot \mathcal{H}s^d_0\cdot U^\dagger+{\rm diag}(V_{eff},0,0),
\ee
with
\be
\mathcal{H}^d_0=\frac{1}{2E_\nu}diag(-\Delta m^2_{21},0,\Delta m^2_{32}).
\ee
Here $U$ is the three neutrino mixing matrix  \citep[i.e. see;][]{gon11}.  From the resonance condition, the effective potential and the oscillation length become
{\small
\be\label{reso3}
V_{eff}-5\times 10^{-7}\frac{\Delta m^2_{32,eV}}{E_{\nu,MeV}}\,\cos2\theta_{13}=0\,\,\,\,\,\,\,{\rm and}\,\,\,\,\,\,\,l_{\rm res}=\frac{4\pi E_{\nu}}{\Delta m^2_{32}\sin 2\theta_{13}}\,,
\ee
}
respectively. The adiabatic condition at the resonance can be expressed as 
{\small
\be
\kappa_{res}\equiv  \frac{2}{\pi}
\left ( \frac{\Delta m^2_{32}}{2 E_\nu} \sin 2\theta_{13}\right )^2
\left (\frac{dV_{eff}}{dr}\right)^{-1} \ge 1\,.
\label{adbcon}
\ee
}
Details are shown in \cite{2014MNRAS.437.2187F}. \\
\\
The neutrino trajectories inside the Earth will depend on the direction of the detector relative to RX J0822-4300 in Puppis A. Neutrinos arriving to the detector will oscillate depending on the nadir angle and hence the matter density that will have to go through in their paths. \\ 
\subsection{Neutrino Expectation}
In general, the neutrino spectrum is computed through the neutrino luminosity  $dN/dE_{\bar{\nu}_e}=  L_{\bar{\nu}_e}/(E^2_{\bar{\nu}_e} 4\pi d^2_z  )$ and used to estimate the neutrino events  
\be\label{rate}
N_{ev}=V N_A\,  \rho_N \int_{T_{\rm burst}} \int_{E'_{\bar{\nu}_e}} \sigma^{\bar{\nu}_ep}_{cc} \frac{dN}{dE_{\bar{\nu}_e}}\,dE_{\bar{\nu}_e}\, dt\,.
\ee

\noindent  The terms $V$ is the fiducial  volume of the neutrino detector,   $\rho_N=2/18\, {\rm g\, cm^{-3}}$ is the density of targets \citep{2004mnpa.book.....M}, $ \sigma^{\bar{\nu}_ep}_{cc}\simeq 9\times 10^{-44}\,\frac{E^2_{\bar{\nu}_e}}{MeV^2}$  is the cross section \citep{1989neas.book.....B} and $dt$ is the time interval of the neutrino burst. The expected events are
\bary\label{num_Neu}
N_{ev}&\simeq&\frac{T_{\rm burst}}{<E_{\bar{\nu}_e}>}V N_A\,  \rho_N  \sigma^{\bar{\nu}_ep}_{cc} <E_{\bar{\nu}_e}>^2<\frac{dN}{ dE_{\bar{\nu}_e}}>\cr
&\simeq&\frac{T_{\rm burst}}{4\pi d^2_z <E_{\bar{\nu}_e}>}V N_A\,  \rho_N  \sigma^{\bar{\nu}_ep}_{cc}\,L_{\bar{\nu}_e}\,.
\eary
Here,  the average energy  $<E_{\bar{\nu}_e}>\simeq E_{\bar{\nu}_e}$ during a period of  time $T_{\rm burst}$ has been assumed \citep{2004mnpa.book.....M}. As follows a brief introduction of the neutrino experiments used in this work is given.
\paragraph{Super-Kamiokande Experiment.} Super-Kamiokande is a Cherenkov detector located at Kamioka in Gifu Prefecture, Japan. It consists of a cylindrical stainless tank whose dimensions are 39 meters in diameter and 42 meters in height, containing 50 kton of ultra-pure water. According to its structure, Super-Kamiokande is divided into an inner and an outer detector region. Currently, the first detector has 11,129 PMTs whereas the outer one has 1885 PMTs. Details  about the technical specifications can be found in \cite{fuk03}.

\paragraph{Hyper-Kamiokande Experiment}   Hyper Kamiokande will be an underground detector located inside a mine in Kamioka Japan. It will have a total fiducial mass of 0.56 millions metric tons of ultra-pure water, which represents a volume 25 times greater than its predecessor Super-Kamiokande. It will also have a symmetric array of 99,000 PMTs in order to detect the Cherenkov light produced during the charged lepton interactions with water nucleons. The main target of this detector will be a more accurate measurement of the neutrino oscillation parameters as well as testing the CP violation in the leptonic sector \citep{abe11, 2014arXiv1412.4673H}.

\paragraph{DUNE Experiment}  The DUNE (\textit{Deep Underground Neutrino Experiment}) experiment will be a neutrino detector with a total fiducial mass of 40 kton of liquid argon located at South Dakota in USA. It will consist of four cryostats instrumented with Liquid Argon Time Projection Chambers (LArTPCs).  This experiment will  begin operating by the year 2026 and among all its main goals are the search for proton decay, more accurate parameter measurements of neutrinos produced in particle accelerators, as well as detection of astrophysical neutrinos. In particular, DUNE will be able to detect neutrinos of transient events such as SNe, Gamma-ray burst and Active Galactic Nuclei, for a better understanding of the dynamics in this kind of sources \citep{acc16}.\\
\\
\section{Results}\label{sec-discussion}
\subsection{Submergence and evolution of the B-field}
\begin{figure*}
\centering
\includegraphics[width=1.\textwidth]{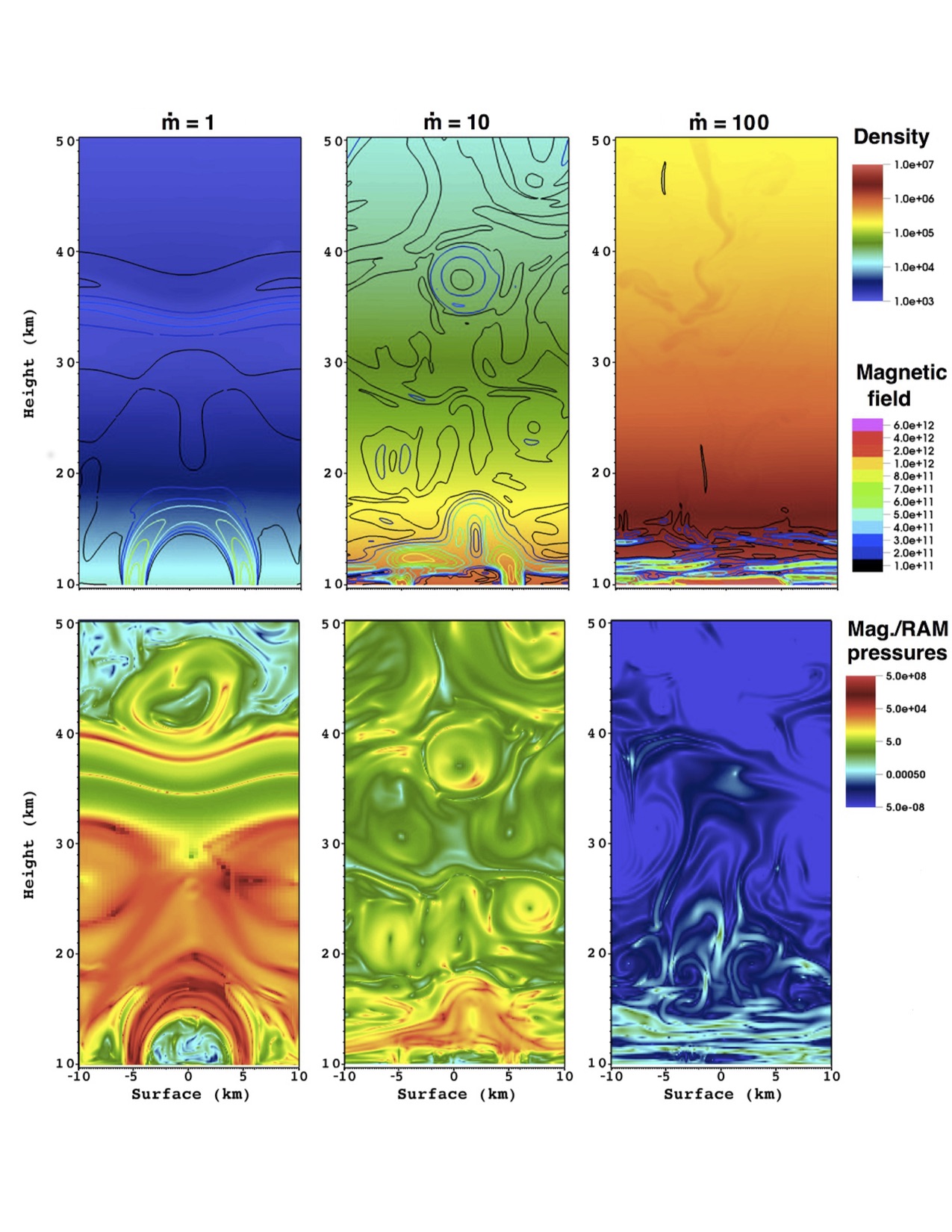}
\caption{Color-maps of density (top panel) and the ratio of the magnetic and ram pressures (bottom panel), with iso-contours of B-field superimposed, for three different accretion rates at $t+200$ ms. The B-field submergence is more evident in the highest accretion rate for this timescale. The magnetic loop resist the hyper-accretion for the lower accretion rate considered, but eventually is compressed and submerged into the stellar crust.}
\label{fig2:simulations}
\end{figure*}

\begin{figure*}
\centering
\includegraphics[width=1.\textwidth]{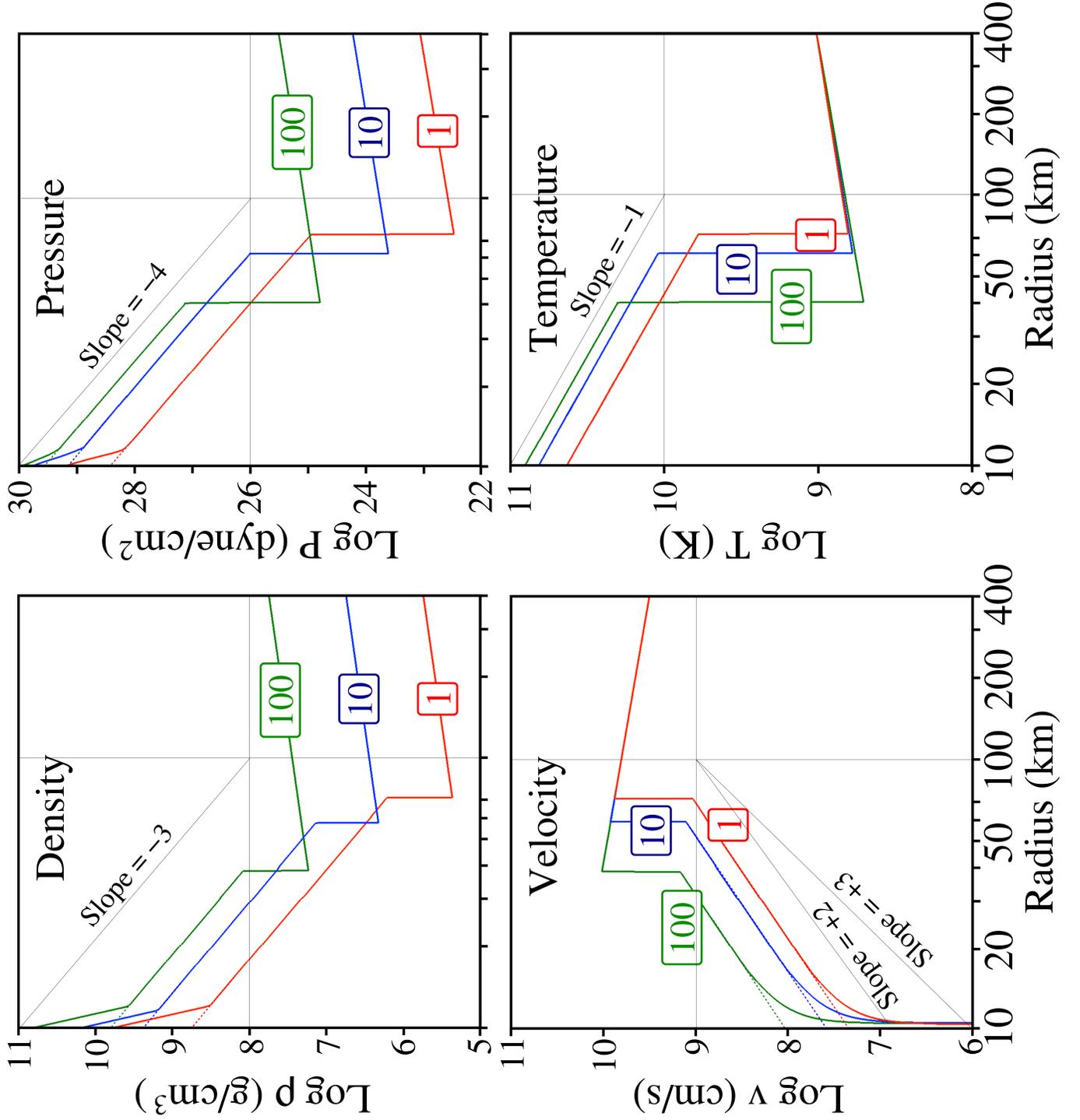}
\caption{Final state radial profiles of density, pressure, absolute velocity and temperature, when the system is completely relaxed and the hydrostatic envelope is well established. The slopes in the curves are the exponents of the analytical approach. The accretion shock and the new crust are evident.}
\label{fig3:profiles}
\end{figure*}

\begin{figure*}
\centering
\includegraphics[width=1.\textwidth]{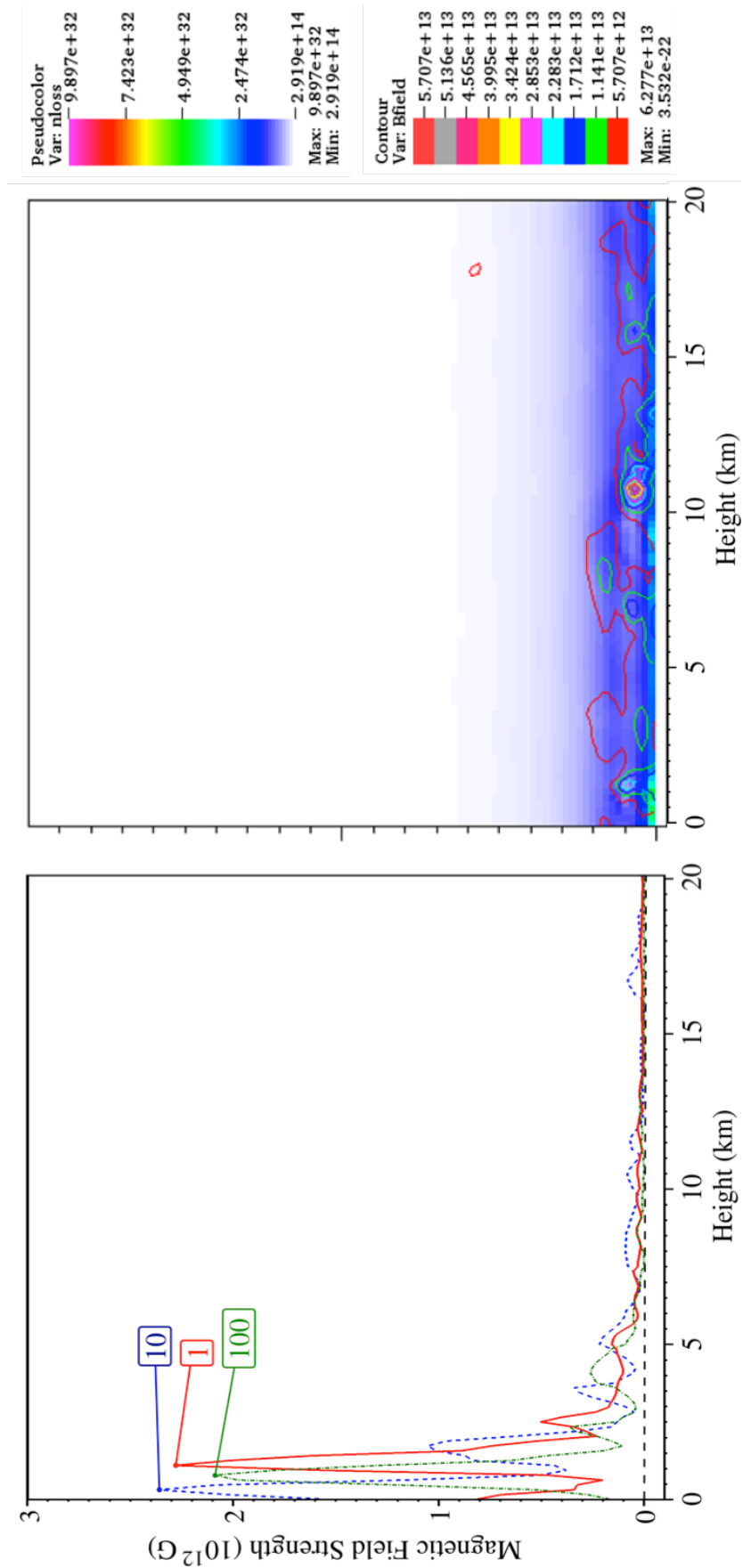}
\caption{Amplified B-field due the high compression inside the neutrino-sphere, for the three accretion rate considered here (left). Close-up of the neutrino emissivity whit iso-contour of B-field superimposed (right). Note the compression of the B-field in the same height-scale of the neutrino-sphere.
}
\label{fig4:B_amp}
\end{figure*}

\begin{figure*}
\centering
\includegraphics[width=0.8\textwidth]{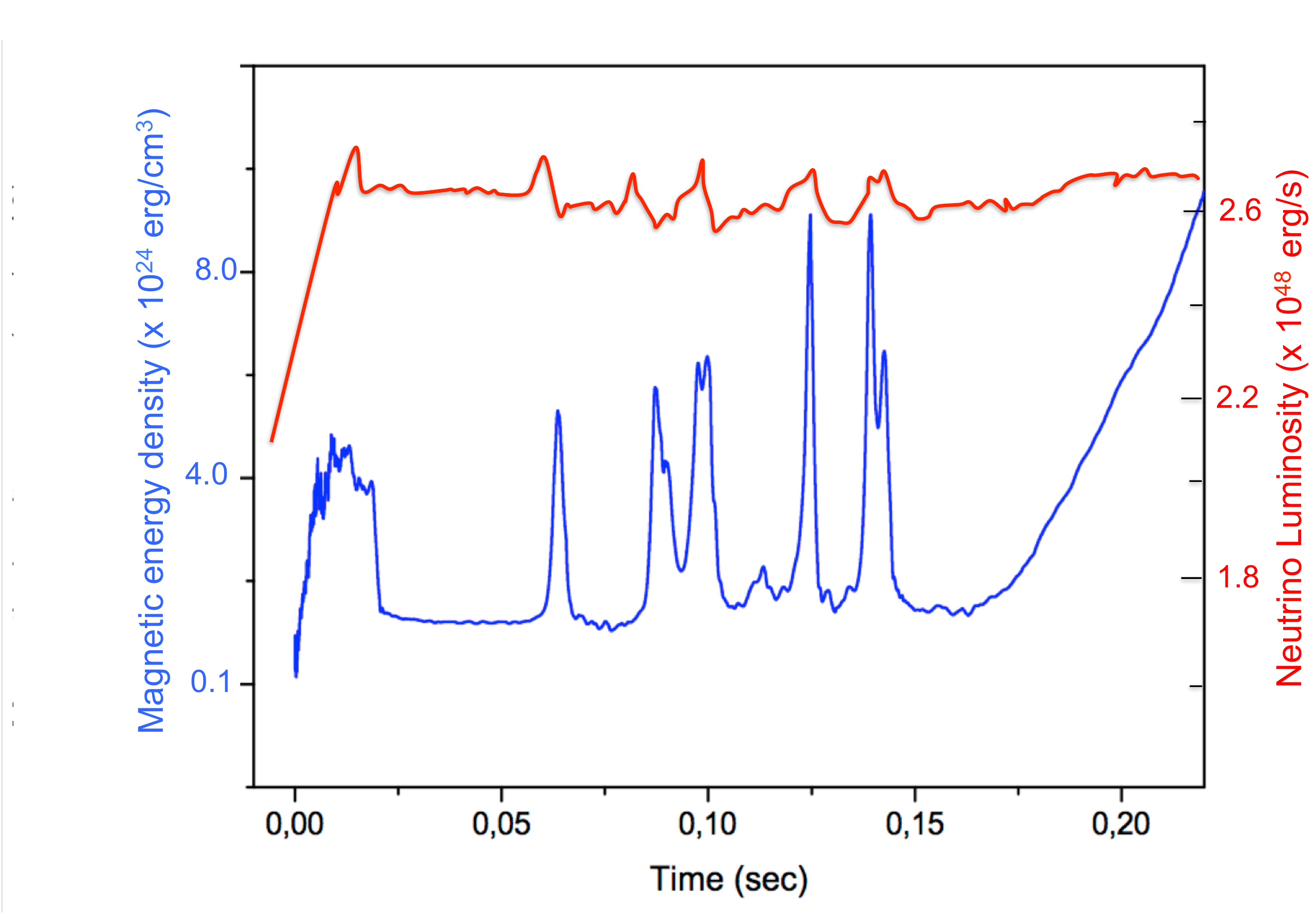}
\caption{Evolution of magnetic energy density (blue) and neutrino luminosity (red)  integrated in the entire computational domain.  Neutrino cooling processes are described in \citep{Itoh1996}.}
\label{fig5:potential}
\end{figure*}

Figure \ref{fig2:simulations} shows colormaps of the plasma density (top panel) for three accretion rates ($\dot{m} = 1, 10, 100 \; M_\odot \, \mathrm{yr}^{-1}$), superimposed with the B-field contours, at $t=200$ ms. Additionally, the ratio between the magnetic pressure and the ram pressure is showed in the bottom panel for the same timescale.  Note that only the highest accretion rate reaches the complete submergence of the B-field in this timescale, as expected.  Although the lower accretion rates take longer to reach this state, it is reached  at $t \sim 600$ ms for $\dot{m} = 10 \; M_\odot \, \mathrm{yr}^{-1}$ and at $t \sim 900$ ms for $\dot{m} = 100 \; M_\odot \, \mathrm{yr}^{-1}$. During the initial transient,  the expansive shock interacts with the flow in free fall forming the accretion shock and a new stellar crust (the place where B-field is dragged and confined).    For the lower accretion rate considered here, the magnetic Rayleigh-Taylor instabilities induce various magnetic reconnection processes between the magnetic loop and a magnetized layer formed above it. The non-magnetized flow falling on the stellar surface encounters in its way the magnetic layer. These interactions induce the formation of  large and isolated fingers which in turn reach the magnetic loop, producing  reconnection processes several times.\\
\\
Figure \ref{fig3:profiles} shows the final radial profiles of the density, the pressure, the velocity and temperature for the values of the accretion rates considered and when the system is relaxed a few hundreds of milliseconds later.   In the final state of the simulation, an accretion envelope in quasi-hydrostatic equilibrium is well established and separated from the continuous accretion inflow by an accretion shock. The material, piled up by the hyper-accretion, builds a new stellar crust, submerging the B-field.  Our numerical results show both the  formation of the stellar crust where the B-field is submerged and the region where neutrinos are created (the so-called neutrino-sphere).\\
\\
Figure \ref{fig4:B_amp} shows the magnitude of the B-field (left) for the accretion rates considered here. We note that the strength of B-field was amplified due to the strong compression inside the crust.   To highlight the neutrino-sphere size, a colormap of the neutrino emissivity with the B-field contours for $\dot{m} = 100 \; M_\odot \, \mathrm{yr}^{-1}$ is shown in the right-hand panel.    The adiabatic and radiative gradients can be estimated through simulations when the system has reached the quasi-hydrostatic equilibrium.  The adiabatic gradient is  $\nabla _{ad}=\frac{\gamma _{c}-1}{\gamma _{c}}\simeq 0.27$ and the radiative gradient is $\nabla _{rad}= \frac{d\ln T}{d\ln P} \simeq 0.26$. Note that the value of the adiabatic index $\gamma _{c}=1.35$ has been taken directly from the simulation, and the radiative gradient was calculated using the temperature and pressure profiles.  It is worth noting that  the gradients are almost constant within the envelope, except in the region near the NS surface.   Due to $\nabla _{rad}\lesssim \nabla _{ad}$, the system is manifestly stable to convection, although being these values close enough numerically they would indicate probably a marginal stability.   After the flow goes through the front of the accretion shock (the sound speed is $c_{s}=\sqrt{\gamma _{c}P/\rho }\simeq 7.3\times 10^{9}\,\mathrm{cm}\ \mathrm{s}^{-1}$ and  $v\simeq 1.3\times 10^{8}\,\mathrm{cm}\,\mathrm{s}^{-1}$ for a Mach number of $m=v/c_{s}\simeq 10^{-2}$), it is fully subsonic inside envelope.  Therefore, the global structure of the accretion column can be studied in detail and compared with the analytical approach presented in \cite{1989ApJ...346..847C}, particularly in the region where the approximations  break down.    It is important to highlight that the thermodynamical conditions in the fluid depend on the accretion towards the proto-NS. The Fermi temperature obtained from the Fermi energy is $T_{F}\simeq 6.5\times 10^{10}\,{\rm K}$.  The temperature obtained from the simulation, close to the bottom of the accretion column in a quasi-stationary state is $T\simeq 4.6\times 10^{10}\,\mathrm{K}$, and then,  $T/T_{F}\simeq 0.7$.  It is thus clear that the assumption of the $e^{\pm}$ degeneracy is not a proper approximation, and a full expression such as the one in the Helmholtz equation of state is required if one wishes to compute the evolution of the flow accurately.  It is also clear that the neutrino cooling effectively turns on a scale of $r\simeq4\times10^{5}\:\mathrm{cm}$.   For the current simulation, the mean value of emissivity in such region is $\dot{\epsilon}_{\nu}\simeq1.6\times10^{30}\:\mathrm{erg\, s^{-1}\, cm^{-3}}$ and the integrated neutrino luminosity is  $L_{\nu}\simeq 3\times10^{48}\:\mathrm{erg\, s^{-1}}$, as indicated in Figure \ref{fig5:potential}.\\
\\
\noindent Figure \ref{fig5:potential} shows the integrated neutrino luminosity (red) and the magnetic energy density (blue) integrated into the whole computational domain for the lower accretion rate considered. Several pronounced peaks are observed on the energy curve.  Each peak corresponds to different magnetic reconnection episodes suffered by the loop (due to interactions between the magnetic Rayleigh-Taylor instabilities and the loop).  Magnetic energy is converted into thermal energy by the code. Also, we show the neutrino luminosity (red) due to several neutrino processes present in the neutrino-sphere of the newly born NS. After an initial transient phase, the neutrino luminosity has small oscillations about a fixed value, with some small peaks, corresponding to the magnetic reconnection episodes suffered by the loop. Magnetic reconnection with very strong  B-fields, as the current case, is a problem poorly understood to date, hence there is not a robust theory that can sustain the results.  The most important scale of the B-field used to compare the result of our analysis is the critical B-field  ($B_{c}$) with the corresponding magnetic energy density given by $U_{c}\simeq7.7\times10^{25}$ erg $\mathrm{cm}^{-3}$. In the present scenario, the maximum  B-field achieved is one order of magnitude less than $B_{c}$. \\
\vspace{0.5cm}
\subsection{Neutrino Oscilations}
\begin{figure*}
\centering
\includegraphics[width=0.7\textwidth]{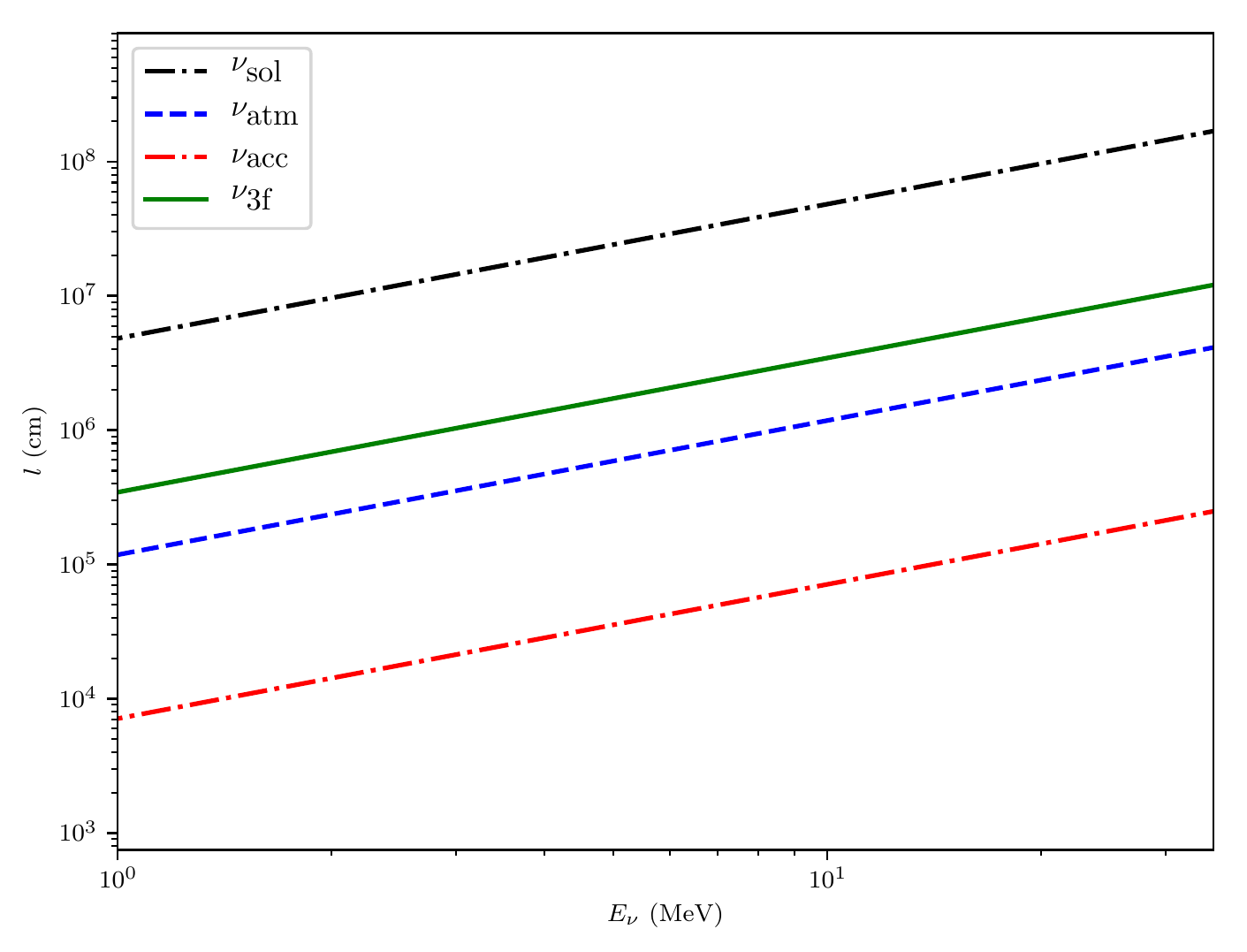}
\caption{Resonance length as a function of neutrino energy considering the two- (Solar, Atmospheric and reactor parameters) and three-flavor mixing.}
\label{fig6:reson_length}
\end{figure*}

\begin{figure*}
\centering
\includegraphics[width=1.\textwidth]{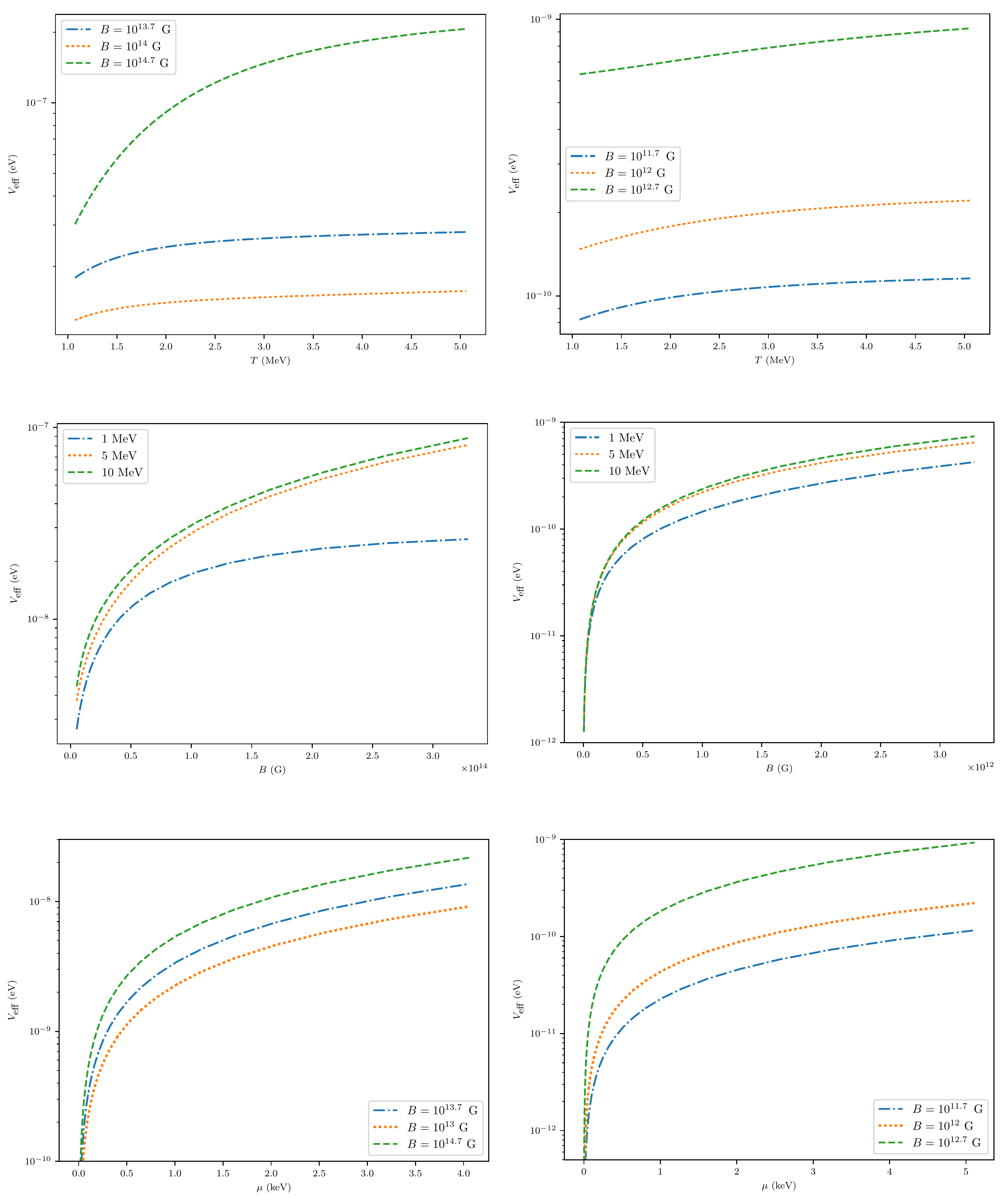}
\caption{Neutrino effective potential as a function of Temperature (top), B-field (medium) and Chemical potential (bottom) at the moderate (left) and weak (right) limits.}
\label{fig6:potentialEff}
\end{figure*}

\begin{figure*}
\centering
\includegraphics[width=1\textwidth]{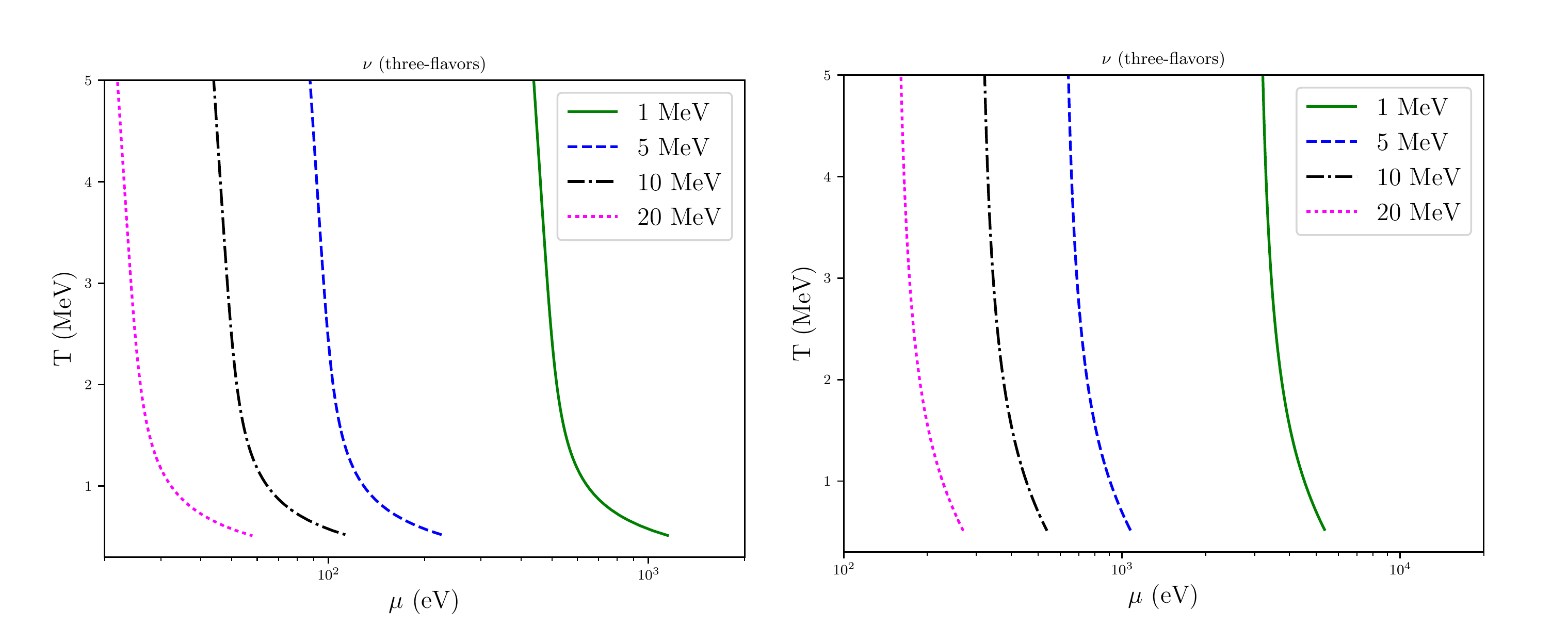}
\caption{Contour lines of Temperature, Chemical potential and Neutrino energy for which the resonance condition is satisfied. For the resonance condition we have used  the effective potential in weak (left) and moderate (right) magnetic limit and the three-flavor parameters.}
\label{fig8:reson_condition}
\end{figure*}

\begin{figure*}
\centering
\includegraphics[width=1.\textwidth]{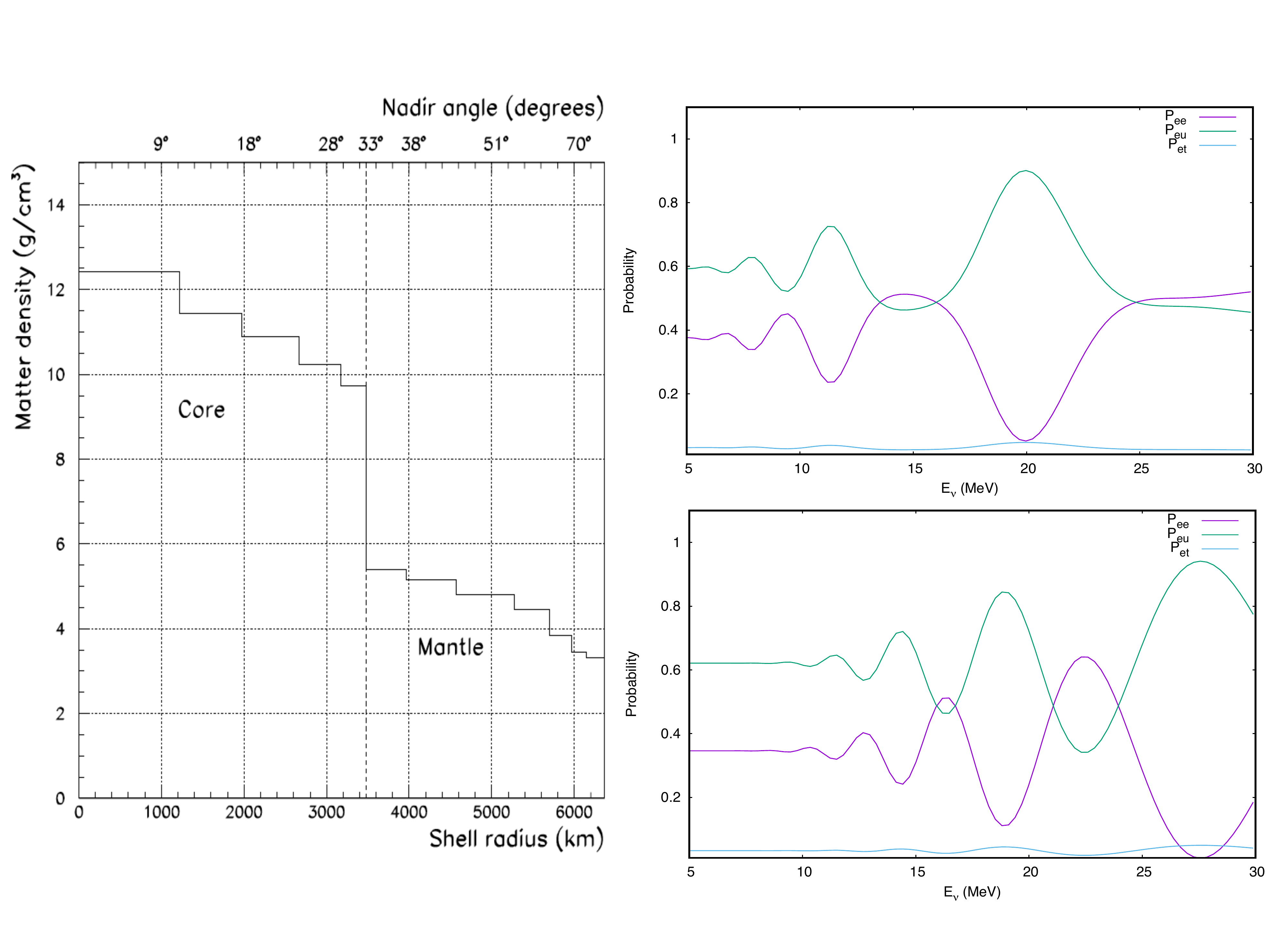}
\caption{Left: Matter density inside the Earth as a function of radius and Nadir angle is shown. Right: Neutrino oscillation probability  as a function of neutrino energy for Nadir angle $\theta=20$ (above) and $\theta=50$ (below).}
\label{fig9:earth}
\end{figure*}

\noindent Figure \ref{fig6:reson_length} shows the resonance length as a function of neutrino energy considering the two- (solar, atmospheric and reactor parameters) and three-flavor mixing. It exhibits that the resonance lengths lie in the range from $5\times 10^3\,$ to $1.8\times 10^7\,$ cm, which is comparable with the size of the new crust on NS surface.  This  figure shows that depending on the oscillation parameters measured in the neutrino experiments, thermal neutrinos can oscillate resonantly before leaving the new crust. For instance, regarding a radius of the new crust $1.3\times 10^6$ cm, only neutrinos with energies less than 5 MeV can oscillate resonantly for atmospheric, accelerator, and three-neutrino mixing parameters but not for solar ones. \\
\\
Figure \ref{fig6:potentialEff} shows the neutrino effective potential in the moderate and weak B-field limits. The left-hand panels display the neutrino effective potential in the moderate B-field limit  as a function of temperature (top panel), B-field (medium panel) and chemical potential (bottom panel).  The effective neutrino potential lies between $\sim 10^{-10}$ and $2.3\times 10^{-7}$ eV for a temperature, B-field and chemical potential in the ranges 1 - 5 MeV,  $10^{13}$ - $4 \times  10^{14}$ G and  0.1 - 4 keV, respectively. It is evident that effective potential increases when the B-field and neutrino energy increases, but not when the temperature increases.  The right-hand panels  exhibit the neutrino effective potential in the weak B-field limit as a function of temperature (top panel),  B-field (medium panel) and chemical potential (bottom panel).  The effective neutrino potential lies between $7 \times 10^{-13}$ and $10^{-9}$ eV for a temperature, B-field and chemical potential in the ranges 1 - 5 MeV,  $10^{11}$ - $4 \times  10^{12}$ G and  0.1 - 5 keV, respectively.  The effective potential in the weak B-field limit has a similar behavior than the moderate one but with different ranges of values.   The effective potential of only neutrino background is not considered hereafter because it is much less than the effective neutrino potential in the moderate and weak B-field limits.\\
\\
For a full description of resonant neutrino oscillations at the new crust of NS, we consider the observable quantities obtained in the simulations during the hypercritical accretion phase.  Taking into account the resonant condition (eq.  \ref{reso3}), we have obtained the range of values of temperature and chemical potential for which this condition is satisfied.  Figure \ref{fig8:reson_condition} shows the contour lines of temperature, chemical potential and neutrino energy for which the resonance condition is satisfied when the effective potentials in moderate (right) and weak (left) B-field limit are considered.   Both panels display that  as neutrino energy decreases the chemical increases when the temperature ranges from 1 to 5 MeV.  The contour lines in both panels represent similar behaviors although with different ranges of values.\\
\\
Figure \ref{fig9:earth} (left) shows  the density of matter inside the Earth as a function of radius and nadir angle. The effective potential is $V_{\rm eff, Ear}=\sqrt2 G_F\, N_A \,Y_e\,\rho_{\rm E}(\Theta)$
where $\rho_{\rm E}(\Theta)$ is the matter density on Earth with $\Theta$ the nadir angle (see Figure \ref{fig9:earth}).  Neutrino oscillation probability  as a function of neutrino energy is shown in Figure \ref{fig9:earth} for $\Theta=20^\circ$ (above) and $\Theta=50^\circ$ (below).  These probabilities are computed using the description shown in \cite{2004SHEP...19...35C}  and \cite{2007APh....27..254A}. In the right-hand panel can be seen that depending on the neutrino's path inside Earth, these will oscillate different up to reach the detector.\\
\\
\\
Thermal neutrinos created during the hypercritical accretion phase will oscillate in matter due to electron density in regions I, II, III and IV,  in vacuum and inside Earth.   In the new crust of NS surface (region I), the plasma  thermalized at $\simeq$ 1 - 4 MeV is submerged in a B-field of $\simeq$ 2.6$\times 10^{10}$ G. Regarding the neutrino effective potential given in this region,  the effective potential for propagating neutrinos  in a thermal medium is plotted.  This figure shows  the positivity of the effective potential ($V_{eff}>$ 0), hence neutrinos can oscillate resonantly.   Using the three- neutrino mixing parameters we study  the resonance condition.   \citet{2014MNRAS.442..239F} calculated  the survival and conversion probabilities for the active-active ($\nu_{e,\mu,\tau} \leftrightarrow \nu_{e,\mu,\tau}$) neutrino oscillations in regions II, III and IV,  showing that neutrinos can oscillate resonantly in these regions.     Taking into consideration the oscillation probabilities in each region and in the vacuum,  we compute the flavor ratio for $E_{\nu}=$1, 5, 15 and 20 MeV.   Table \ref{Table} displays a small deviation from the standard ratio of  flavor 1:1:1.   In this calculation, we take into account the neutrino cooling processes of $e^-e^+$ pair annihilation, Plasmon decay and  Photo-neutrino emission.  Taking into consideration that the effective potential for propagating neutrinos are larger than zero,  we have used in it,  neutrinos instead of anti-neutrinos.\\
\begin{table*}[h!]      
\begin{center}
\caption[]{The neutrino flavor ratio on the surface of each zone.}\label{Table}
\begin{minipage}{126mm}
\begin{tabular}{lccccccccc}
\hline
 
 $E_{\nu}$  & On the NS surface & Accretion material & Free fall zone& Outer layers & On Earth \\
 \\\hline \hline 

{\small 1}     &  {\small 1.17:0.92:0.91}  &  {\small 1.181:0.905:	0.914} & {\small 1.149:0.928:0.923} &  {\small 1.115:0.942:0.943}  &  {\small 1.031:0.991:0.978} \\
\\

{\small 5}   & {\small 1.18:0.91:0.91}  &  {\small 1.158:0.919:0.923}&  {\small 1.162:0.919:0.919} &  {\small 1.156:0.922: 0.922}  &  {\small 1.045:0.984:0.971}\\
\\

{\small 15}  & {\small 1.19:0.91:0.90} &  {\small 1.160:0.920:0.920} &  {\small 1.142:0.929:0.929} &  {\small 1.139:0.933:0.928}  &  {\small 1.037:0.988:0.975}   \\
\\

{\small 20}  &  {\small 1.2:0.90:0.90}  &   {\small 1.159: 0.921:0.920}  &   {\small 1.149:0.926:0.925} &  {\small 1.125:0.941:0.934}  &  {\small 1.035:0.991:0.974}  \\\hline
 
\end{tabular}
\end{minipage}
\end{center}
\end{table*}
\vspace{1cm}
%
%
\subsection{Expected Neutrino Events}
\begin{figure*}
\centering
\includegraphics[width=\textwidth]{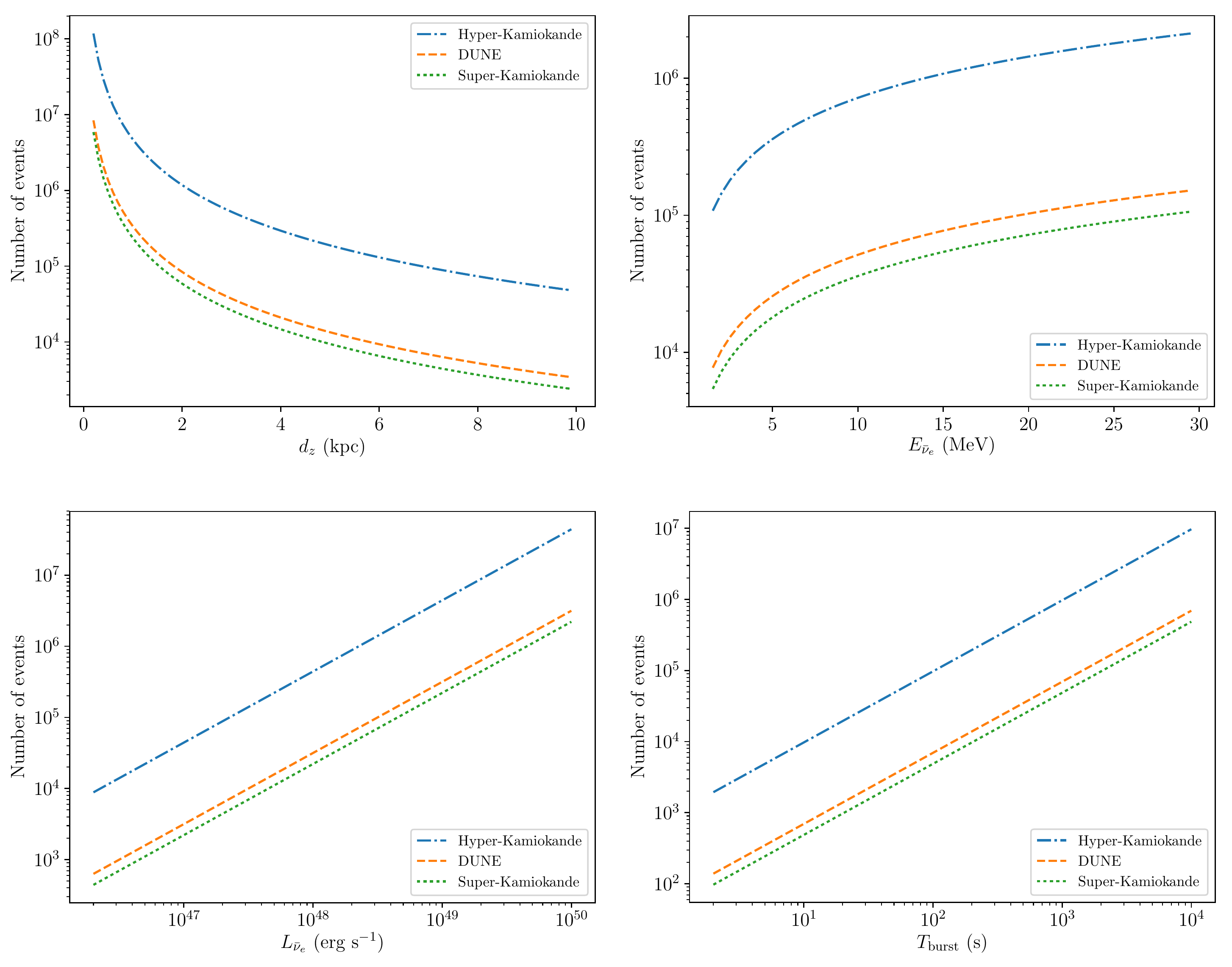}
\caption{Number of neutrino events expected on Super-Kamiokande (dotted green line), Hyper-Kamiokande (dotted-dashed blue line) and DUNE (dashed orange line) experiments are considered.   The upper left-hand panel corresponds to the number of events as a function of distance for $L_{\bar{\nu}_e}=2.6\times10^{48}\:\mathrm{erg\, s^{-1}}$, $ T_{\rm burst}=10^3\, {\rm s}$ and $E_{\bar\nu_e} = 13.5\,{\rm MeV}$.   The upper right-hand panel shows the number of events as a function of the average neutrino energy  for $L_{\bar{\nu}_e}=2.6\times10^{48}\:\mathrm{erg\, s^{-1}}$, $T_{\rm burst}=10^3\, {\rm s}$ and $d_z =2.2\,{\rm kpc}$.  The lower left-hand panel exhibits the number of events as a function of the neutrino luminosity for $d_z =2.2\,{\rm kpc}$, $ T_{\rm burst}=10^3\, {\rm s}$ and $E_{\bar\nu_e} = 13.5\,{\rm MeV}$.   The lower right-hand panel displays the number of events as a function of  the duration of the neutrino burst    for $L_{\bar{\nu}_e}=2.6\times10^{48}\:\mathrm{erg\, s^{-1}}$, $d_z =2.2\,{\rm kpc}$ and $E_{\bar\nu_e} = 13.5\,{\rm MeV}$.}
\label{fig10:events}
\end{figure*}

Using ranges of typical values for the distance $0.1 \lesssim d_z \lesssim 10\,{\rm kpc}$,   the average neutrino energy  $1 \lesssim E_{\bar\nu_e} \lesssim 30\,{\rm MeV}$,  the neutrino luminosity  $10^{46} \lesssim L_{\bar\nu_e} \lesssim 10^{50}\,{\rm erg\, s^{-1}}$ and the duration of the neutrino burst $1 \lesssim T_{\rm burst} \lesssim 10^4\,{\rm s}$, the number of events expected in the SK (dotted green line), HK (dotted-dashed blue line) and DUNE (dashed orange line) experiments from the hypercritical phase are plotted in Figure \ref{fig10:events}.   The upper left-hand panel corresponds to the number of events as a function of distance for $L_{\bar{\nu}_e}=2.6\times10^{48}\:\mathrm{erg\, s^{-1}}$, $ T_{\rm burst}=10^3\, {\rm s}$ and $E_{\bar\nu_e} = 13.5\,{\rm MeV}$. This panel shows that the neutrino events lie in the range of $2\times 10^3$ and $1.3\times 10^8$.  It can be inferred that neutrino events from the hypercritical phase in CCO candidates \citep{2010PNAS..107.7147K} such as Kesteven 79, Cassiopeia A, G353.6-0.7,  PKS 1209-51/52 and Puppis A located at 7.1 kpc, 3.4 kpc, 2.2 kpc, 4.5 kpc and 2.2 kpc respectively, could be expected in the SK, HK and DUNE experiments \citep{2018arXiv180907057F}. The upper right-hand panel shows the number of events as a function of the average neutrino energy  for $L_{\bar{\nu}_e}=2.6\times10^{48}\:\mathrm{erg\, s^{-1}}$, $T_{\rm burst}=10^3\, {\rm s}$ and $d_z =2.2\,{\rm kpc}$.  This panel shows that the neutrino events range from $5\times 10^3$ to $2.6\times 10^6$.  The lower left-hand panel exhibits the number of events as a function of the neutrino luminosity for $d_z =2.2\,{\rm kpc}$, $ T_{\rm burst}=10^3\, {\rm s}$ and $E_{\bar\nu_e} = 13.5\,{\rm MeV}$. This panel shows that the neutrino events lie in the range of $4\times 10^2$ and $5.2\times 10^7$. The lower right-hand panel displays the number of events as a function of  the duration of the neutrino burst    for $L_{\bar{\nu}_e}=2.6\times10^{48}\:\mathrm{erg\, s^{-1}}$, $d_z =2.2\,{\rm kpc}$ and $E_{\bar\nu_e} = 13.5\,{\rm MeV}$. This panel shows that the neutrino events range from $93$ to $1.1\times 10^7$ which corresponds to the worst ($93$) and the best ($1.1\times 10^7$) scenery for SK and HK, respectively.
\\
\\
The values of observable quantities, such as the distance $2.2\pm 0.3$ kpc \citep{1995AJ....110..318R}, the neutrino luminosity  $L_{\bar{\nu}_e}=(2.6\pm 0.2)\times10^{48}\:\mathrm{erg\, s^{-1}}$,  the average neutrino energy  $E_{\bar\nu_e} = 13.5\,{\rm MeV}$ \citep{2007fnpa.book.....G}, were used considering the results obtained in our simulations, the location of Puppis A and the values reported for SN1987A. It can be observed that average the MeV neutrinos from the hypercritical accretion phase in CCO candidates with similar characteristics like Puppis A can be detected in the new generation of neutrino experiments.\\
\\
We estimate the number of the initial neutrino burst from the NS formation. Taking into consideration  the typical duration of the neutrino burst $t\simeq10$ s,   T$\approx$ 4 MeV  \citep{2007fnpa.book.....G}, the average neutrino energy $<E_{\bar{\nu}_e}>\simeq  13.5\pm 3.2\, {\rm  MeV}$ \citep{2007fnpa.book.....G} and the total fluence equivalent for Puppis A  $\Phi\approx(1.30\pm 0.54) \times 10^{12}\, \bar{\nu}_e \,{\rm cm^{-2}}$,  the total number of neutrinos released and the total radiated luminosity during the NS formation are $N_{tot}=6\,\Phi\, 4\pi\, d_z^2\approx(4.55\pm 1.82) \times 10^{57}$ and  $L_\nu\approx \frac{N_{tot}}{t}\times <E_{\bar{\nu}_e}>\approx (9.84 \pm3.94) \times 10^{51}$ erg/s, respectively. Finally, the number of events expected during the NS formation on HK experiment is $N_{ev}\simeq (4.82\pm 1.91)\times 10^6$. Therefore,  we realize that  during the NS formation are expected much more events ($\sim 10^4  - 10^5$ events) than the  hyper-accretion phase.\\
%
%
%
\section{Conclusions}\label{sec-Results}
We have studied the dynamic of hyper-accretion onto the newly born NS in RX J0822-4300.  It was performed using numerical 2D-MHD simulations through the AMR FLASH method. We have found that for the hyper-accretion rate considered here, initially the B-field resists to be submerged in the new crust suffering several episodes of magnetic reconnections and returning to its original shape. This is very interesting because it allows us to analyze how the magnetic reconnection processes work in extreme conditions of temperature, density and strong B-fields.   Finally, the B-field is submerged into de new crust by the hyper-accretion, with a quasi-hydrostatic envelope around it.  In addition, we have estimated the neutrino luminosity considering the entire involved processes which are mandatory for the analysis of propagation/oscillations of thermal neutrinos through the distinct regions of the pre-supernova.    We have observed that  several peaks in the neutrino luminosity curve are very notorious. The same behavior is exhibited in the curve of magnetic energy density. This corresponds to several episodes of magnetic reconnections driven by magnetic Rayleigh-Taylor instabilities.  The code allows us to capture the rich morphology and the relevant characteristics of this phase such as the neutrino losses at the base of the envelope among others.   While neutrino losses play an important role in the envelope structure, the optical depth of photons is large enough so that the radiative transfer only plays a role in the later evolution. The escape of radiation at the Eddington limit can have a dramatic effect on the evolution of the hyper-accretion process.   A detailed study of the thermal structure of the gas surrounding the NS is needed to determine whether it takes place and also to investigate if the quasi-hydrostatic envelope is evaporated by neutrinos, photons or some other process.   Although further numerical studies of these phenomena are necessary, the results shown in this paper provide a glimpse of the complex dynamics around the newly born NS, moments after the core-collapse supernova explosion.\\
\\
In this work, we have required the reverse shock model to explain the fallback of material onto the compact remnant.  Following \cite{1989ApJ...346..847C}  we have performed MHD simulations and found that our results not only  recover the radial profiles predicted by this model for the quasi-hydrostatic envelope structure, but also allow us to follow the evolution of the system during the formation of the new crust on the stellar surface with the consequent submergence of the B-field.  On the other hand,  the fallback mechanism was revisited by \cite{2014arXiv1401.3032W}, in the core-collapse SN context. They discussed the current mechanisms proposed in the literature to explain the fallback in the early history of SN: the rarefaction wave deceleration, the energy and momentum loss of the ejecta, and the reverse shock deceleration.   Using several numerical codes to carry out simulations in high dimensionality, the authors found that, for two particular progenitor models, the first two mechanisms called ``prompt fallback mechanisms" would produce fallback at early times more efficiently (of the order of 15 s) than the reverse shock mechanism.  In this case it must wait for the forward shock to reach the hydrogen layer (typically a few hundred seconds) before the reverse shock is produced, leading to fallback that takes place a few hundred to a thousand seconds after the launch of the explosion. As a future project, it would be interesting to perform new numerical simulations and redo our accounts, using the prompt fallback mechanism, to perform a comparative analysis with our present results.\\
\\
We have done a full description of emission, propagation, and oscillations of thermal neutrinos created during the hypercritical accretion phase. We have considered neutrinos with energies ranging from 1 to 30 MeV and cooling processes of   $e^-e^+$ pair annihilation, Plasmon decay and  Photo-neutrino emission.  We have used the neutrino self-energy and the neutrino effective potential up to order $m_W^{-4}$ as a function of temperature, chemical potential, B-field and neutrino energy at the moderate and weak B-field limits. We have computed the neutrino oscillations on the NS surface, the accretion material, the free fall zone, the outer layer, the vacuum between the compact object RX J0822-4300 and Earth and finally, inside Earth.\\   
The resonance lengths as a function of neutrino energy for the two- (solar, atmospheric and reactor parameters) and three-flavor mixing were computed. We found that the resonance lengths range from $5\times 10^3$ to $1.8\times 10^7$ cm, which are comparable with the new crust of NS surface.\\
The effective potential as a function of  temperature, B-field and chemical potential at the moderate and weak B-field limited was analyzed. We found that the effective potential lies in the range from $1.3\times 10^{-10}$ to $2.3\times 10^{-7}$ eV  for the moderate B-field limit and $7\times 10^{-13}$ to $1.1\times 10^{-9}$ eV for the weak B-field limit.\\
A description of resonant neutrino oscillations for three-neutrino mixing at the new crust of NS was performed. We found that the chemical potential lies in the range of  $20$ to $1.1\times 10^3$ eV and  $180$ to $5.8\times 10^3$ eV for the effective potential in the moderate and weak B-field limits, respectively, for a temperature between 1 to 5 MeV.\\
We have estimated the number of neutrino events and the standard flavor ratio. The neutrino events expected from the hyper-accretion phase on Super-Kamiokande, Hyper-Kamiokande and DUNE experiments range from 5 to 2500 events which exhibit a nonsignificant deviation of the standard flavor ratio (1:1:1).  Neutrinos of energies ranging from 1 to 30 MeV are similar to those produced by a SN type II.   These Galactic events make the thermal neutrino flux high on Earth compared with the neutrino burst seen from SN 1987A. \\
\acknowledgements
We thank  Dany Page Rollinet and John Beacom  for useful discussions. NF acknowledge financial  support  from UNAM-DGAPA-PAPIIT  through  grant  IA102917. 
%

\begin{thebibliography}{}
\expandafter\ifx\csname natexlab\endcsname\relax\def\natexlab#1{#1}\fi

\bibitem[{{Abe} {et~al.}(2018){Abe}, {Bronner}, {Haga}, {Hayato}, {Ikeda},
  {Iyogi}, \& et~al.}]{2018PhRvD..97g2001A}
{Abe}, K., {Bronner}, C., {Haga}, Y., {et~al.} 2018, \prd, 97, 072001

\bibitem[{Abe {et~al.}(2011)Abe, Abe, Aihara, Fukuda, Hayato, Huang, Ichikawa,
  Ikeda, Inoue, Ishino, {et~al.}}]{abe11}
Abe, K., Abe, T., Aihara, H., {et~al.} 2011, arXiv preprint arXiv:1109.3262

\bibitem[{{Abe} {et~al.}(2017){Abe}, {Amey}, {Andreopoulos}, {Antonova},
  {Aoki}, {Ariga}, {Ashida}, {Autiero}, {Ban}, {Barbi}, \&
  et~al.}]{2017arXiv170406409A}
{Abe}, K., {Amey}, J., {Andreopoulos}, C., {et~al.} 2017, ArXiv e-prints,
  arXiv:1704.06409

\bibitem[{{Abe} {et~al.}(2012){Abe}, {Aberle}, {dos Anjos}, {Barriere},
  {Bergevin}, {Bernstein}, \& et~al.}]{2012PhRvD..86e2008A}
{Abe}, Y., {Aberle}, C., {dos Anjos}, J.~C., {et~al.} 2012, \prd, 86, 052008

\bibitem[{Acciarri {et~al.}(2016)Acciarri, Acero, Adamowski, Adams, Adamson,
  Adhikari, Ahmad, Albright, Alion, Amador, {et~al.}}]{acc16}
Acciarri, R., Acero, M., Adamowski, M., {et~al.} 2016, arXiv preprint
  arXiv:1601.02984

\bibitem[{{Adri{\'a}n-Mart{\'{\i}}nez}
  {et~al.}(2012){Adri{\'a}n-Mart{\'{\i}}nez}, {Al Samarai}, {Albert},
  {Andr{\'e}}, {Anghinolfi}, {Anton}, \& et~al.}]{2012PhLB..714..224A}
{Adri{\'a}n-Mart{\'{\i}}nez}, S., {Al Samarai}, I., {Albert}, A., {et~al.}
  2012, Physics Letters B, 714, 224

\bibitem[{{Agafonova} {et~al.}(2007){Agafonova}, {Aglietta}, {Antonioli},
  {Bari}, {Boyarkin}, {Bruno}, {Fulgione}, {Galeotti}, {Garbini}, {Ghia},
  {Giusti}, {Kemp}, {Kuznetsov}, {Kuznetsov}, {Malguin}, {Menghetti}, {Pesci},
  {Pless}, {Porta}, {Ryasny}, {Ryazhskaya}, {Saavedra}, {Sartorelli}, {Selvi},
  {Vigorito}, {Vissani}, {Votano}, {Yakushev}, {Zatsepin}, \&
  {Zichichi}}]{2007APh....27..254A}
{Agafonova}, N.~Y., {Aglietta}, M., {Antonioli}, P., {et~al.} 2007,
  Astroparticle Physics, 27, 254

\bibitem[{{Bahcall}(1989)}]{1989neas.book.....B}
{Bahcall}, J.~N. 1989, {Neutrino astrophysics}

\bibitem[{{Bernal} \& {Fraija}(2016)}]{2016MNRAS.462.3646B}
{Bernal}, C.~G., \& {Fraija}, N. 2016, \mnras, 462, 3646

\bibitem[{{Bernal} {et~al.}(2010){Bernal}, {Lee}, \& {Page}}]{Bernal2010}
{Bernal}, C.~G., {Lee}, W.~H., \& {Page}, D. 2010, \rmxaa, 46, 309

\bibitem[{{Bernal} {et~al.}(2013){Bernal}, {Page}, \& {Lee}}]{Bernal2013}
{Bernal}, C.~G., {Page}, D., \& {Lee}, W.~H. 2013, \apj, 770, 106

\bibitem[{{Cavanna} {et~al.}(2004){Cavanna}, {Costantini}, {Palamara}, \&
  {Vissani}}]{2004SHEP...19...35C}
{Cavanna}, F., {Costantini}, M.~L., {Palamara}, O., \& {Vissani}, F. 2004,
  Surveys in High Energy Physics, 19, 35

\bibitem[{{Chevalier}(1989)}]{1989ApJ...346..847C}
{Chevalier}, R.~A. 1989, \apj, 346, 847

\bibitem[{{Chevalier}(2005)}]{Chevalier2005}
---. 2005, \apj, 619, 839

\bibitem[{{de Salas} {et~al.}(2017){de Salas}, {Forero}, {Ternes}, {Tortola},
  \& {Valle}}]{2017arXiv170801186D}
{de Salas}, P.~F., {Forero}, D.~V., {Ternes}, C.~A., {Tortola}, M., \& {Valle},
  J.~W.~F. 2017, ArXiv e-prints, arXiv:1708.01186

\bibitem[{{Fraija}(2014{\natexlab{a}})}]{2014MNRAS.437.2187F}
{Fraija}, N. 2014{\natexlab{a}}, \mnras, 437, 2187

\bibitem[{{Fraija}(2014{\natexlab{b}})}]{2014ApJ...787..140F}
---. 2014{\natexlab{b}}, \apj, 787, 140

\bibitem[{{Fraija} \& {Bernal}(2015)}]{2015MNRAS.451..455F}
{Fraija}, N., \& {Bernal}, C.~G. 2015, \mnras, 451, 455

\bibitem[{{Fraija} {et~al.}(2014){Fraija}, {Bernal}, \&
  {Hidalgo-Gam{\'e}z}}]{2014MNRAS.442..239F}
{Fraija}, N., {Bernal}, C.~G., \& {Hidalgo-Gam{\'e}z}, A.~M. 2014, \mnras, 442,
  239

\bibitem[{{Fraija} {et~al.}(2018){Fraija}, {Bernal}, {Morales}, \&
  {Negreiros}}]{2018arXiv180907057F}
{Fraija}, N., {Bernal}, C.~G., {Morales}, G., \& {Negreiros}, R. 2018, ArXiv
  e-prints, arXiv:1809.07057

\bibitem[{{Fryxell} {et~al.}(2000){Fryxell}, {Olson}, {Ricker}, {Timmes},
  {Zingale}, {Lamb}, {MacNeice}, {Rosner}, {Truran}, \& {Tufo}}]{Fryxell2000}
{Fryxell}, B., {Olson}, K., {Ricker}, P., {et~al.} 2000, \apjs, 131, 273

\bibitem[{Fukuda {et~al.}(2003)Fukuda, Fukuda, Hayakawa, Ichihara, Ishitsuka,
  Itow, Kajita, Kameda, Kaneyuki, Kasuga, {et~al.}}]{fuk03}
Fukuda, S., Fukuda, Y., Hayakawa, T., {et~al.} 2003, Nuclear Instruments and
  Methods in Physics Research Section A: Accelerators, Spectrometers, Detectors
  and Associated Equipment, 501, 418

\bibitem[{{Geppert} {et~al.}(1999){Geppert}, {Page}, \&
  {Zannias}}]{1999A26A...345..847G}
{Geppert}, U., {Page}, D., \& {Zannias}, T. 1999, \aap, 345, 847

\bibitem[{{Giacconi} {et~al.}(1979){Giacconi}, {Branduardi}, {Briel},
  {Epstein}, {Fabricant}, {Feigelson}, {Forman}, {Gorenstein}, {Grindlay},
  {Gursky}, \& et~al.}]{1979ApJ...230..540G}
{Giacconi}, R., {Branduardi}, G., {Briel}, U., {et~al.} 1979, \apj, 230, 540

\bibitem[{{Giunti} \& {Chung}(2007)}]{2007fnpa.book.....G}
{Giunti}, C., \& {Chung}, W.~K. 2007, {Fundamentals of Neutrino Physics and
  Astrophysics} (Oxford University Press)

\bibitem[{{Gonzalez-Garcia}(2011)}]{gon11}
{Gonzalez-Garcia}, M.~C. 2011, Physics of Particles and Nuclei, 42, 577

\bibitem[{{Goriely} \& {Janka}(2016)}]{2016MNRAS.459.4174G}
{Goriely}, S., \& {Janka}, H.-T. 2016, \mnras, 459, 4174

\bibitem[{{Gotthelf} {et~al.}(2013){Gotthelf}, {Halpern}, \&
  {Alford}}]{2013ApJ...765...58G}
{Gotthelf}, E.~V., {Halpern}, J.~P., \& {Alford}, J. 2013, \apj, 765, 58

\bibitem[{{Hui} \& {Becker}(2006)}]{2006A&A...454..543H}
{Hui}, C.~Y., \& {Becker}, W. 2006, \aap, 454, 543

\bibitem[{{Hyper-Kamiokande Working Group} {et~al.}(2014){Hyper-Kamiokande
  Working Group}, {:}, {Abe}, {Aihara}, {Andreopoulos}, {Anghel}, {Ariga},
  {Ariga}, {Asfandiyarov}, {Askins}, \& et~al.}]{2014arXiv1412.4673H}
{Hyper-Kamiokande Working Group}, {:}, {Abe}, K., {et~al.} 2014, ArXiv
  e-prints, arXiv:1412.4673

\bibitem[{{IceCube Collaboration} {et~al.}(2014){IceCube Collaboration},
  {Aartsen}, {Ackermann}, {Adams}, {Aguilar}, {Ahlers}, {Ahrens}, {Altmann},
  {Anderson}, {Arguelles}, \& et~al.}]{2014arXiv1410.7227I}
{IceCube Collaboration}, {Aartsen}, M.~G., {Ackermann}, M., {et~al.} 2014,
  ArXiv e-prints, arXiv:1410.7227

\bibitem[{{Itoh} {et~al.}(1996){Itoh}, {Hayashi}, {Nishikawa}, \&
  {Kohyama}}]{Itoh1996}
{Itoh}, N., {Hayashi}, H., {Nishikawa}, A., \& {Kohyama}, Y. 1996, \apjs, 102,
  411

\bibitem[{{Kaspi}(2010)}]{2010PNAS..107.7147K}
{Kaspi}, V.~M. 2010, Proceedings of the National Academy of Science, 107, 7147

\bibitem[{{Korobkin} {et~al.}(2012){Korobkin}, {Rosswog}, {Arcones}, \&
  {Winteler}}]{2012MNRAS.426.1940K}
{Korobkin}, O., {Rosswog}, S., {Arcones}, A., \& {Winteler}, C. 2012, \mnras,
  426, 1940

\bibitem[{Lee(2013)}]{LeeDongwook2013}
Lee, D. 2013, J. Comput. Phys., 243, 269

\bibitem[{{Mohapatra} \& {Pal}(2004)}]{2004mnpa.book.....M}
{Mohapatra}, R.~N., \& {Pal}, P.~B. 2004, {Massive neutrinos in physics and
  astrophysics}

\bibitem[{{Muslimov} \& {Page}(1995)}]{1995ApJ...440L..77M}
{Muslimov}, A., \& {Page}, D. 1995, \apjl, 440, L77

\bibitem[{{Pac}(2018)}]{2018arXiv180104049P}
{Pac}, M.~Y. 2018, ArXiv e-prints, arXiv:1801.04049

\bibitem[{{Pavlov} {et~al.}(2004){Pavlov}, {Sanwal}, \&
  {Teter}}]{2004IAUS..218..239P}
{Pavlov}, G.~G., {Sanwal}, D., \& {Teter}, M.~A. 2004, in IAU Symposium, Vol.
  218, Young Neutron Stars and Their Environments, ed. F.~{Camilo} \& B.~M.
  {Gaensler}, 239

\bibitem[{{Petre} {et~al.}(1996){Petre}, {Becker}, \&
  {Winkler}}]{1996ApJ...465L..43P}
{Petre}, R., {Becker}, C.~M., \& {Winkler}, P.~F. 1996, \apjl, 465, L43

\bibitem[{{Popov} {et~al.}(2015){Popov}, {Kaurov}, \& {Kaminker}}]{Popov2015}
{Popov}, S.~B., {Kaurov}, A.~A., \& {Kaminker}, A.~D. 2015, \pasa, 32, 18

\bibitem[{{Reynoso} {et~al.}(1995){Reynoso}, {Dubner}, {Goss}, \&
  {Arnal}}]{1995AJ....110..318R}
{Reynoso}, E.~M., {Dubner}, G.~M., {Goss}, W.~M., \& {Arnal}, E.~M. 1995, \aj,
  110, 318

\bibitem[{{Sahu} {et~al.}(2009){Sahu}, {Fraija}, \&
  {Keum}}]{2009JCAP...11..024S}
{Sahu}, S., {Fraija}, N., \& {Keum}, Y.-Y. 2009, \jcap, 11, 24

\bibitem[{{Seo}(2017)}]{2017arXiv171008204S}
{Seo}, S.-H. 2017, ArXiv e-prints, arXiv:1710.08204

\bibitem[{{Shabaltas} \& {Lai}(2012)}]{Shabaltas2012}
{Shabaltas}, N., \& {Lai}, D. 2012, \apj, 748, 148

\bibitem[{{T2K Collaboration} {et~al.}(2017){T2K Collaboration}, {Abe}, {Amey},
  {Andreopoulos}, {Antonova}, {Aoki}, {Ariga}, {Autiero}, {Ban}, {Barbi}, \&
  et~al.}]{2017arXiv170100432T}
{T2K Collaboration}, {Abe}, K., {Amey}, J., {et~al.} 2017, ArXiv e-prints,
  arXiv:1701.00432

\bibitem[{{The NOvA Collaboration} {et~al.}(2017{\natexlab{a}}){The NOvA
  Collaboration}, {Adamson}, {Aliaga}, {Ambrose}, {Anfimov}, {Antoshkin}, \&
  et~al.}]{2017arXiv170303328T}
{The NOvA Collaboration}, {Adamson}, P., {Aliaga}, L., {et~al.}
  2017{\natexlab{a}}, ArXiv e-prints, arXiv:1703.03328

\bibitem[{{The NOvA Collaboration} {et~al.}(2017{\natexlab{b}}){The NOvA
  Collaboration}, {Adamson}, {Aliaga}, {Ambrose}, {Anfimov}, {Antoshkin}, \&
  et~al.}]{2017arXiv170105891T}
---. 2017{\natexlab{b}}, ArXiv e-prints, arXiv:1701.05891

\bibitem[{{Timmes} \& {Swesty}(2000)}]{Timmes2000}
{Timmes}, F.~X., \& {Swesty}, F.~D. 2000, \apjs, 126, 501

\bibitem[{{Vigan{\`o}} {et~al.}(2013){Vigan{\`o}}, {Rea}, {Pons}, {Perna},
  {Aguilera}, \& {Miralles}}]{2013MNRAS.434..123V}
{Vigan{\`o}}, D., {Rea}, N., {Pons}, J.~A., {et~al.} 2013, \mnras, 434, 123

\bibitem[{{Wong} {et~al.}(2014){Wong}, {Fryer}, {Ellinger}, {Rockefeller}, \&
  {Kalogera}}]{2014arXiv1401.3032W}
{Wong}, T.-W., {Fryer}, C.~L., {Ellinger}, C.~I., {Rockefeller}, G., \&
  {Kalogera}, V. 2014, ArXiv e-prints, arXiv:1401.3032

\end{thebibliography}
%

%
\clearpage


\end{document}